\documentclass[prl,aps,showpacs,
preprint,
nofootinbib,
floatfix,superscriptaddress]{revtex4}
\usepackage{graphicx}
\usepackage{epsfig}

\newcommand{\beq}{\begin{equation}}
\newcommand{\eeq}{\end{equation}}
\newcommand{\beqs}{\begin{eqnarray}}
\newcommand{\eeqs}{\end{eqnarray}}

\newcommand{\Tr}{{\rm Tr}}

\def\hbar{\hspace{0pt}\raisebox{1pt}{$-$} \hspace{-7pt} h}

\def\di{\mbox{d}}

\begin{document}
\title{Precision electro-weak parameters
 from $AdS_5$, localized kinetic terms  and anomalous dimensions.}

\author{Maurizio Piai \thanks{email:piai@u.washington.edu }}
\affiliation{Department of Physics, University of Washington,
Seattle, WA 98195}

\date{August 21, 2006}

%\vspace{6mm}

\begin{abstract}

I compare the tree level estimate of the electro-weak precision  parameters in 
two (exactly solvable) toy models of dynamical symmetry breaking 
in which the strong dynamics is assumed to be described by
a five-dimensional (weakly coupled) gravity dual.
I discuss the effect of brane-localized kinetic terms,
their use as regulators for the couplings of otherwise non-normalizable modes,
and the impact of a large deviation from its natural value
for the scaling dimension of the background field responsible for spontaneous
symmetry breaking. 
The latter is assumed to model the effects of {\it walking} dynamics,
i.e. of a large anomalous dimension of the chiral condensate,
it has a strong impact of the spectrum of 
spin-1 fields and, as a consequence,  on the electro-weak precision parameters.
The main conclusion is that models of dynamical symmetry breaking
based on a large-$N_c$ strongly interacting
$SU(N_c)$ gauge theory are compatible with 
precision electro-weak constraints, and produce
a very distinctive signature testable at the LHC. 
Some of the considerations discussed are directly relevant
for analogous models in the context of $AdS-QCD$.

\end{abstract}


\maketitle

%%%%%%%%%%%%%%%%%%%%%%%%%%%%%%%%%%%%%%%%%%%%%%%%%%%%%%%%%%%%%%%%%%%%%%
\section{Introduction}
%%%%%%%%%%%%%%%%%%%%%%%%%%%%%%%%%%%%%%%%%%%%%%%%%%%%%%%%%%%%%%%%%%%%%%

Some  special super-Yang-Mills conformal
field theories are known to admit a dual description
in terms of a (weakly interacting) higher-dimensional 
gravity theory~\cite{Maldacena} in a negative-curvature background 
($AdS_5$ space). This suggests the 
speculative idea that a much larger
class of strongly interacting gauge theories,
in which not only conformal invariance, but also
supersymmetry are (softly) broken,
might admit such a dual description.
It is hence interesting to explore the space of 
the five-dimensional models that can be obtained 
with simple and controllable deformations of the 
pure $AdS_5$ background,
looking for the gravity-dual of a wider class of strongly interacting
four-dimensional theories, with the hope of learning something
about phenomenologically relevant strongly interacting theories
that would otherwise be difficult to study.

The starting point of the simplest such construction consists of writing the effective action
of a gauge theory in a five-dimensional space-time 
 containing a warped gravity background described by the
metric:
\beqs
\di s^2 &=& \left(\frac{L}{z}\right)^{2}\left( \eta_{\mu\nu}\di x^{\mu}\di x^{\nu}\,-\,\di z^2\right)\,,
\eeqs
where $x^{\mu}$ are four-dimensional coordinates, $\eta_{\mu\nu}$ the Minkoski 
metric with signature $(+,-,-,-)$, and $z$ is the extra (warped) dimension.
The dimensionful parameter $L$ is the $AdS_5$ curvature, and sets the 
overall scale of the model.
Schematically, the interpretation in terms of four dimensional 
conformal theory relates the  
rescaling in the fifth dimension $z$ to conformal transformations
in the four-dimensional dual description.
Conformal symmetry is broken by  the boundaries 
\beqs
L_0\,<\,z\,<\,L_1\,,
\eeqs
with $L_0>L$, where $L_0$ and $L_1$ correspond to the UV and IR cut-offs of the
conformal theory.
The gauge symmetry of the five-dimensional bulk is related
to the global symmetries of the dual CFT.
More details about the general construction and interpretation of these models
can be found elsewhere in the rich literature on the subject  
(see for instance~\cite{pheno} for a simple, clear 
and general summary of the basic elements of these constructions).

Very recently, this approach has been use 
in order to formulate an Effective Field Theory (EFT) description of 
QCD at the energies above the 
range of validity of the chiral Lagrangian~\cite{pomarol}\cite{AdSQCD}\cite{Strassler},
in order to give a simple description of the physics of 
(strongly interacting) mesonic resonances (see also~\cite{karch}~\cite{others}).
Besides possible modifications of the gravity background and
of the field content of the EFT,  these models differ
by the 
assumed bulk profile of the chiral symmetry breaking background,
by the  introduction of  dilaton-type backgrounds, and by 
how the UV and IR cut-offs are introduce and regulated.

Besides QCD, another class of strongly interacting theories 
relevant for phenomenology contains
the models of dynamical electro-weak symmetry breaking,
generically referred to as technicolor (TC)~\cite{TC}.
Apart from the very different energy scale, the non-linear sigma model
 description of TC  is very similar
 to the chiral Lagrangian of QCD, the main difference being
 given by the fact that in the former a subset of the chiral 
 symmetry is (weakly)  gauged and corresponds to the Standard Model 
 $SU(2)_L\times U(1)_Y$ gauge group.
 It is hence natural  to construct a five-dimensional
 weakly coupled gravity model with the appropriate symmetry content so
 as to describe the physics responsible for electro-weak symmetry breaking,
 aimed at the study of 
 the EFT in the energy window ranging from the $W$-boson mass 
 to a few orders of magnitude above the electro-weak scale itself.
 Part of this range will be soon explored at the LHC, and it is hence crucial
 to have models that are compatible with all present data, but
 produce new distinctive signatures accessible at these higher energies.
 
Models in this class are strongly constrained by experimental data on 
precision electro-weak observables~\cite{Peskin}~\cite{Barbieri},
in particular on the parameters $\hat{S}$ and $\hat{T}$ to be defined later on.
Based on dimensional analysis,
one expects them to scale as the ratio of the
electro-weak gauge boson masses and mass differences 
to the mass of the lightest
spin-1 excited state of the strong sector ({\it techni-$\rho$}) as
$
\hat{S}  \propto M_W^2/M_{\rho}^2$ and 
$\hat{T}  \propto  (M_Z^2-M_W^2)/{M_{\rho}^2}$
up to multiplicative  model-dependent factors determined by the strong TC dynamics.
These estimates are, generically speaking, 
too large in comparison with experimental data,
unless the mass of the techni-$\rho$ is pushed unnaturally 
far above the electro-weak scale, in the multi-TeV range.
It would be impossible to detect these new states directly even at the LHC.

However, a non-trivial departure from QCD-like behavior of the
underlying TC-dynamics might stabilize a substantial hierarchy between different scales,
and even  produce significant  suppression factors
in the computation of $\hat{S}$ and $\hat{T}$.
Such a possibility is exemplified by
TC models with walking behaviour~\cite{walking}:
the presence of a regime in which the theory is quasi conformal, 
together with the large anomalous dimension of the chiral condensate,
can change in a substantial way the dependence of the masses
of fermions and gauge bosons as a function of the 
(dynamically generated) scales in the underlying theory,
and affect the precision electro-weak parameters
(see for instance~\cite{AS}).

The effect of walking behavior  on the phenomenology of SM fermions
has been studied extensively, and is known to be very big.
In TC, fermion masses 
are introduced by coupling two techni-quark fields and  two SM fermions via
a four-fermion operator, with a (dimensionful) coefficient
$1/\Lambda_{ETC}^2$~\cite{ETC}. $\Lambda_{ETC}$ is the scale, much higher than
the electro-weak scale, at which the (global) flavor symmetries of the SM are broken.
Walking modifies the dependence on this scale of the 
mass of the SM fermions $m_f$. With no big anomalous dimensions,
$m_f \propto \Lambda_{TC}^3/\Lambda_{ETC}^2$, where $\Lambda_{TC}$
is in the range of the electro-weak scale. 
In full generality, this scaling 
is dictated by the dimension $d$ of the condensate, and given by
\beqs
m_f &\propto& \Lambda_{TC}\left(\frac{\Lambda_{TC}}{\Lambda_{ETC}}\right)^{d-1}\,,
\eeqs
in such a way as to gain an enhancement factor of the order of a power of $\Lambda_{ETC}/\Lambda_{TC}$ if $d<3$.
This enhancement factor is important
for obtaining large enough masses for the fermions while at the same time suppressing
Flavor Changing Neutral Currents (FCNC). 

Walking behavior is also expected to affect the phenomenology
of the spin-1 states of the theory. Many studies have been carried on
in the literature in order to gauge the magnitude of such effect,
which is generally believed to reduce via, non-perturbative effects,
 the perturbative estimates of precision electro-weak parameters.
The major obstacle to a precise refinement of this statements is
the fact that it is very difficult to compute reliably the precision 
parameters $\hat{S}$ and $\hat{T}$ in the context of
a four-dimensional $SU(N_c)$ strongly interacting gauge theory~\cite{Shrock}. 

 In this paper, inspired by the works on $AdS-QCD$,
 I discuss some examples of the use of the techniques
 developed there for the construction of models of 
 dynamical electro-weak symmetry breaking.  The main
 interest of my analysis is to compute 
 the precision observables $\hat{S}$ and $\hat{T}$,
 and to study how the results depend on the assumptions
 used in constructing the effective Lagrangian in five-dimensions.
 Previous studies can be found in the literature of 
 the Higgsless models~\cite{higgsless}, in the context of 
 composite Higgs models~\cite{composite} and deconstruction~\cite{deconstruction}.
 Most of the results on the strong-dynamics 
 rely ultimately on the idea of 
 vector-dominance and hidden local symmetry~\cite{hidden},
 and could be rewritten in term
 of four-dimensional deconstructed models~\cite{deconstructQCD},
 but the number of free parameters is far smaller
 in the $AdS-CFT$ context, and manifest conformal symmetry plays
 a crucial role in the present study.
 
 I  study a set of five-dimensional models with a
 $SU(2)_L\times U(1)_Y$ gauge symmetry in the bulk,
 the lightest  modes of which correspond to the 
 photon, and to the standard-model $W$ and $Z$ gauge bosons.
 Electro-weak symmetry breaking is induced by the background 
 vacuum expectation value (VEV) of a bulk scalar field,
(the  dual description of the chiral condensate).
The (four-dimensional) physical  mass of the scalar excitations 
is assumed to be large, so that 
the non-linear sigma-model description applies.

 I do not include the standard-model fermions. 
 Ordinary quarks and leptons are fundamental fields, because they are not
 supposed to carry TC interactions, and hence are
 confined to live on the UV brane~\footnote{A possible exception could be represented by the top quark. Models in which the dynamics of the third generation has a special role have been studied at length in order to explain the largeness of the top mass~\cite{top}. See however~\cite{APS1}.}. The effect of their existence
 is reflected here in the introduction of (divergent) localized kinetic terms.
  These break explicitly the conformal
 invariance, and provide a natural regulator for the theory~\cite{HR}.
 In the absence of such boundary terms, in the limit in which
 one sends to infinity the UV cut-off, the zero-modes of the gauge fields
 become non-normalizable and decouple from the spectrum.
 The regulator allows to take the limit of infinite cut-off while keeping the gauge coupling of
 the zero modes finite. In this way, the final EFT depends only on 
 quantities that are physically well defined at low energy: 
 the scale of confinement (position of the IR boundary)
 and the value of the couplings of the heavy resonances (gauge coupling in the bulk) 
 and of the electro-weak gauge bosons (gauge couplings on the UV-brane), with
 no explicit dependence on the UV cut-off of the model.

 The present study aims primarily to illustrate the effect on the spectrum and
 on the precision parameters of 
 the conformal symmetry and of  the large anomalous dimension
 of the chiral symmetry breaking condensate, and hence a semi-realistic toy-model 
 is enough to capture the main interesting dynamical features.
 For instance, I do not consider the effect of introducing a custodial symmetry
 in the bulk  to suppress $\hat{T}$, since this would just add to the technical 
 difficulties, without substantial changes to the phenomenology.
 For the same reason, I do not include {\it different} boundary terms for
 (nor a dilaton background with different coupling to)
 the axial and vector components of the gauge fields~\cite{HS}.

 The main part of the analysis consists of the computation of the 
polarization tensors for the SM gauge bosons in the $AdS_5$ background 
 in the cases in which the chiral symmetry breaking condensate has two different
(large)  anomalous dimensions.
 The spectrum of spin-1 states,  as well as the estimate
 of  the precision observables, are significantly modified by the anomalous dimensions. 
 This can be  a useful tool for the construction of
 more realistic, testable models of dynamical electro-weak symmetry breaking.
  
The paper is organized as follows.
I first review the definitions and experimental bounds
on precision electro-weak parameters.
Then I devote two sections to the definition of the models
under consideration and to the algebraic manipulations
that lead to the polarizations. These two sections 
are very detailed, and intended for the reader
who is not familiar with the $AdS-CFT$ language. 
The following two sections are devoted to the 
explicit derivation of the electro-weak precision parameters
in the two models.
Finally, the phenomenology is presented, with comparison 
to the experimental limits, and I conclude with a 
critical discussion and interpretation of the whole procedure.
 
 %%%%%%%%%%%%%%%%%%%%%%%%%%%%%%%%%%%%%%%%%%%%%%%%%%%%%%%%%%%%%%%%%%%%%%
\section{Precision Parameters.}
%%%%%%%%%%%%%%%%%%%%%%%%%%%%%%%%%%%%%%%%%%%%%%%%%%%%%%%%%%%%%%%%%%%%%%
Before entering the specific discussion of the five-dimensional models,
I briefly recall here the basic formulae and experimental constraints from precision 
electro-weak physics.
The (bilinear part of the) 
gauge boson sector of the Standard Model Lagrangian in 4 dimensions can be 
written as (after integrating out all heavy states)
\beqs
{\cal L} &=&\frac{P_{\mu\nu}}{2}A_{i}^{\mu}\pi_{ij}(q^2)A_{j}^{\nu}\,+\,g_{4}^{a}J_{a\mu}A_a^{\mu}\,,
\eeqs
where the index $i$ runs over the $SU(2)_L\times U(1)_Y$ generators,
 $P_{\mu\nu}=\eta_{\mu\nu}-q_{\mu}q_{\nu}/q^2$, the $g_4^a =g_4,g^{\prime}_4$
 are the SM gauge couplings of $SU(2)_L\times U(1)_Y$
 and $\pi_{ij}$ is the polarization
 tensor of the SM gauge bosons.
The precision electro-weak parameters of interest here are  defined by:
\beqs
\hat{S}&\equiv&\frac{g_4}{g_4^{\prime}}\,\pi_{WB}^{\prime}(0)\,,\\
\hat{T}&\equiv&\frac{1}{M_W^2}\left(\pi_{WW}(0)-\pi_{+}(0)\right)\,,
\eeqs
where 
$M_W^2$ is the mass of the $W$-boson, and $\pi^{\prime}=\di \pi/\di q^2$,
and where I call $\pi_{WW}$ and $\pi_{WB}$ and $\pi_{BB}$ the entries of the
$2\times 2$ polarization tensor in the $T_3$ direction, while $\pi_{+}$ is
the polarization in the $T_{1,2}$ direction of $SU(2)_L$. 

I take as indicative of the experimentally allowed ranges
(at the $3\sigma$ level):
\beqs
\hat{S}_{exp}&=& (-0.9\pm 3.9) \times 10^{-3}\,,\\
\hat{T}_{exp}&=& (2.0\pm3.0) \times 10^{-3}\,,
\eeqs
from~\cite{Barbieri}, with the caution that  these are bounds extrapolated to
the case of a Higgs boson with mass of $800$ GeV\footnote{Some of the approximations
used in the extraction of these limits from the experimental data do not rigorously apply
in this range, since the comparison is done with the Standard Model at the one-loop level,
and for large masses of the Higgs boson, i.e. large effective quartic coupling, 
the loop expansion is not well behaved. Also,
the strongly interacting sector I am describing here does not contain a Higgs field at all,
since this is supposed to be very heavy, and is integrated out constructing a
non-linear sigma-model description of symmetry breaking. The dependence on the Higgs mass
should hence be replaced with the dependence on the UV cut-off of the effective 
theory, which is not a physical quantity, and hence introduces 
uncertainty in the procedure.}. 

These bounds, in particular the one on $\hat{S}$,
are much more stringent than the ones obtained including
$Z$-pole observables only~\cite{Barbieri}, often referred to
in the literature. The main
message is that both the observables have to lie in the 
few $\times 10^{-3}$ range, and that a positive $\hat{T}$
is actually favored (at the $1\sigma$ level)
in the case in which there is no light Higgs in the spectrum.
However, the fact that  the comparison is done here in a non rigorous way suggests 
that probably the error bars are under-estimated, and that hence
the results of the study carried on in this paper 
have to be understood as conservative.

The polarizations  $\pi_i$ defined above 
 can be rewritten in terms of the propagator
of the SM gauge fields, defined as the boundary values of the five-dimensional gauge bosons,
i.e. as the fields that couple to the localized SM currents.
The propagators associated with charged currents $i p_{+}$  and 
vectorial and axial-vectorial  neutral currents $i p_{v}$ and $i p_{a}$ are
related to the $\pi_i$ tensors by:
\beqs
\pi_{+}&=&\frac{1}{p_{+}}\,,
\eeqs
and
\begin{widetext}
\beqs
\pi&=&\frac{1}{(g_4^2+g_4^{\prime 2})p_a p_v}\left(\begin{array}{cc}
g_4^2p_v+g_4^{\prime\,2}p_a & -g_4g_4^{\prime}(p_v-p_a)\cr
-g_4g_4^{\prime}(p_v-p_a)& g_4^2p_a+g_4^{\prime\,2}p_v\end{array}\right)\,,
\eeqs 
\end{widetext}
from which
\begin{widetext}
\beqs
\label{Sandpolarizations}
\hat{S}&=&\frac{M_W^2}{M_Z^2}\left(\pi_v-\pi_a\right)^{\prime}(0)\,,\\
\hat{T}&=&\frac{1}{M_W^2}\left(\frac{M_W^2}{M_Z^2}\pi_a-\pi_{+}\right)(0)\,.
\eeqs
\end{widetext}
Here, due to the fact that this is going to be a tree-level analysis of the 
new-physics  contributions to the precision observables, I traded the
(weak) gauge couplings for the masses of the gauge bosons, using
the (tree-level) relations of the Standard Model. The error introduced in
this way is negligible at the present level of precision.

The information about the heavy modes in the theory
will be contained in the corrections to the propagators of
the photon, $W$ and $Z$ gauge bosons (denoted as $p_{v}$, $p_{+}$ and $p_{a}$),
after integrating out the heavy excitations,
and will depend on the masses and coupling constants of the heavy modes.
A crucial assumption I am working with throughout this whole
study is that, at the  EFT level at which none of the heavy resonances 
has been yet integrated out,  the only mixing between light and heavy states 
is present in the mass matrices
for the gauge bosons, neglecting the more general case of a non-trivial kinetic
mixing.

It is worth recalling that the combination of all the precision measurements,
besides those performed at the $Z$ pole, yields significant constraints
on a large set of  universal and non-universal parameters. The latter are mainly
affected by the coupling to fermions, and hence are not discussed here. 
Of the former, only $\hat{S}$ and $\hat{T}$ are directly relevant in the present context,
since the others are produced only by higher-derivative terms in the Taylor expansion
of the polarization tensors, and hence can be neglected.

Finally, before turning to the models, it is useful to look more in details at the
expression of $\hat{S}$ in Eq.~(\ref{Sandpolarizations}).
This can be recast, using dispersion relations, as
\beqs
\label{sSdispersion}
\hat{S} &\propto& g_4^2\sum_n\left(\frac{f_{\rho\,n}^{ 2}}{M_{\rho\,n}^{2}}
-\frac{f_{a_1\,n}^{ 2}}{M_{a_1\,n}^2}\right)\,,
\eeqs
where  $f_{\rho,a_1\,n}$ are
the decay constants of the heavy resonances, and $M_{\rho,a_1\,n}$
their masses. This expression will be derived and discussed better later
on, and for the moment neglects the presence of possible suppression factors.

It is not difficult to see how to reduce, in a generic phenomenological 
model, the contribution to $\hat{S}$, playing with three different 
possibilities. The most naif one is to make the masses of the 
resonances bigger. This could render them too heavy for observation.
A second way is to suppress the decay constants. This can be done by
tuning the coupling of the heavy resonances to the electro-weak currents,
in respect to the coupling of the (lightest) SM gauge bosons.
This tuning is unnatural, because it requires to assume the presence of very a very strong coupling
in the strong sector, which makes impossible to perform any kind of computation.
It also would make the 
resonances more difficult to observe, since at LHC the proton-proton 
initial scattering process can be described in terms of Standard Model currents, and
hence this strategy would reduce the production probability.
Finally, a more appealing possibility is to arrange for a cancellation
between the axial-axial and vector-vector contributions to $\hat{S}$.
This approach~\cite{AS} could lead to light enough new states
(with large enough couplings to the
SM currents)  as to allow for their detection at the LHC.

This third mechanism requires to tune the masses and decay constants
in order for the cancellation to take place.
This is exact in the limit in which there is no isospin violation. 
The goal is hence to construct a model
 in which  the lightest states are directly sensitive to electro-weak symmetry
 breaking, while the heavy resonances are less sensitive to it.
 This is what is expected to happen in models
 with walking dynamics, in which the  (strong) running gauge coupling approaches
 an IR fixed-point, and hence heavy modes are only marginally 
 affected by the blowing off of the coupling itself in the IR and the consequent 
 formation of a symmetry breaking condensate.
 
 This paper aims at a more quantitative discussion of the size
 and model dependence of the aforementioned three mechanisms,
 and at the study of their feasibility within models
 that try to minimize fine-tuning without loosing 
 testability at the LHC and predictive power.

%%%%%%%%%%%%%%%%%%%%%%%%%%%%%%%%%%%%%%%%%%%%%%%%%%%%%%%%%%%%%%%%%%%%%%
\section{The Model(s).}
%%%%%%%%%%%%%%%%%%%%%%%%%%%%%%%%%%%%%%%%%%%%%%%%%%%%%%%%%%%%%%%%%%%%%%
I adopt the conventions for the metric defined in the Introduction.
In particular, the determinant of the metric is $\sqrt{G} = (L/z)^{5}$
for the $AdS_5$ background.

The field content consists of a single complex scalar $\Phi$ transforming 
as a $(2,2)$ of $SU(2)_L\times SU(2)_R$. I gauge the $SU(2)_L\times U(1)_Y$ 
subgroup, in which the generator of $U(1)_Y \subset SU(2)_R$ is the $T_3$, 
with $T_i=\tau_i/2$ and $\tau_i$ the 
Pauli matrices.

The bulk action for $\Phi$ and the gauge bosons $L = L_i T_i$ of $SU(2)_L$  
and $R = R_3 T_3$ of $U(1)_Y$ is given by:
\beqs
{\cal S}_{5} &=& \int\di^4 x \int_{L_0}^{L_1}\di z\,\sqrt{G}\left[\frac{}{}
\Tr\left(G^{MN}D_M\Phi D_N\Phi-M^2|\Phi|^2\right)\,\nonumber\right.\\
&& \left.-\frac{1}{2}\Tr\left(L_{MN}L_{RS}+R_{MN}R_{RS}\right)G^{MR}G^{NS}\right]\,,
\eeqs
and the boundary terms are given by
\beqs
{\cal S}_{4} &=& \int\di^4 x \int_{L_0}^{L_1}\di z \,\sqrt{G}\left[-\frac{1}{2} D\, \delta(z-L_0)\right.\\
&&\Tr\left[L_{\mu\nu}L_{\rho\sigma}+R_{\mu\nu}R_{\rho\sigma}\right]
G^{\mu\rho}G^{\nu\sigma}\nonumber\\
&&+C\,\delta(z-L_0)\Tr\left[G^{\mu\nu}D_{\mu}\Phi D_{\nu}\Phi\right]\nonumber\\
&& -\delta(z-L_0)\,2\lambda_0 \left(\Tr|\Phi|^2-\frac{1}{2}\mbox{v}_0^2\right)^2\nonumber\\
&&\left. -\delta(z-L_1)\,2\lambda_1 \left(\Tr|\Phi|^2-\frac{1}{2}\mbox{v}_1^2\right)^2\right]\nonumber\,,
\eeqs
where the covariant derivative is given by
\beqs
D_M\Phi &=& \partial_M \Phi + i (g L_M \Phi - g^{\prime} \Phi R_M)\,,\nonumber
\eeqs
and where the Yang-Mills action is written in terms of the antisymmetric 
field-strength tensors $L_{\mu\nu}$ and $R_{\mu\nu}$ (for most of the following,
these tensors are approximated 
by neglecting the quadratic terms).
In the action, $M^2$ is a bulk mass term for the scalar, 
and $g$ and $g^{\prime}$ are
the (dimensionful) gauge couplings in five-dimensions.

The choice of boundary terms is dictated by the rules of
holographic renormalization~\cite{HR}: the presence of 
the UV brane, and the assumption that SM fermions are
localized on it, introduces an explicit breaking of
conformal invariance, that would manifest itself with 
(localized and divergent) radiative corrections to the kinetic terms of the 
bulk fields, and hence require the presence of  $C$ and $D$.
I am not going to discuss the complete 
spectrum in this paper,  but just focus on the spin-1 modes of the 4-dimensional
action, treating the vacuum expectation value (VEV) of $\Phi$ 
as a background.

The first step consists of solving the equations of motion for the 
lowest mode of the scalar field~\cite{GW}.
In the limit $\lambda_i \rightarrow +\infty$ ,
 the physical scalar mass diverges. Hence I consider the non-linear realization,
 in which the transverse degrees of freedom are set to zero and
 decoupled.  I use the same notation  $\Phi$ also after
 these massive fluctuations around the VEV are integrated out,
 and hence  I write $\Phi$ in terms of its background value as:
\beqs
\Phi (x,z) & = & \frac{1}{2} \mbox{v}(z)e^{2i \phi(x,z)/\mbox{v}(z) }
\eeqs
in which $\phi=\phi_iT_i$ are the (would-be) Goldstone bosons, and the VEV
is assumed to be constant in each Minkoski slice of the space.
The boundary terms for the scalar potential reduce to the constraints on the 
(classical) background $\mbox{v}(z)$:
\beqs
\mbox{v}(L_0)&=&\mbox{v}_0\,,\\
\mbox{v}(L_1)&=&\mbox{v}_1\,,
\eeqs

I consider two distinct cases in the following, so defined
\begin{itemize}
\item
AdS background with bulk mass term $M^2=-3/L^2$, condensate of dimension $d=1$,
\item
AdS background with  bulk mass term $M^2=-4/L^2$, condensate of dimension $d=2$.
\end{itemize}

With AdS metric the scalar background satisfies:
\beqs
\partial_z \left(\frac{L^3}{z^3}\partial_z\mbox{v}\right) -\frac{L^5}{z^5} M^2 \mbox{v} &=& 0\,,\\
\eeqs
 the general solutions of which depend on the choice of $M^2$ in the following way:
\beqs
\mbox{v} (z)&=& A z + B z ^3 \,,\, \mbox{for} \,M^2=-3/L^2\,,\\
\mbox{v} (z)&=& A z^2 + B z ^2 \log(z/L) \,,\, \mbox{for} \,M^2=-4/L^2\,,
\eeqs
with $A$ and $B$ determined by the boundary conditions.
The special choices for the value of $M^2$ discussed here are  
dictated by  the dictionary of $AdS/CFT$. The symmetry breaking
VEV of the scalar field corresponds to the condensate of the 
operator breaking the chiral symmetry in the four-dimensional dual,
and its scaling in the fifth dimension corresponds to the
scaling dimension of the condensate.

 I choose the boundary terms for $\Phi$ in such a way as
to set $B=0$ at finite $L_0 > L > 0$. This choice reduces to
\beqs
\frac{\mbox{v}_0}{L_0} = \frac{\mbox{v}_1}{L_1}\,,\,\mbox{for} \,M^2=-\frac{3}{L^2}\,,\\
\frac{\mbox{v}_0}{L^2_0} = \frac{\mbox{v}_1}{L^2_1}\,,\,\mbox{for} \,M^2=-\frac{4}{L^2}\,,
\eeqs
which yields
\beqs
\mbox{v} (z)&=& \frac{\mbox{v}_1}{L_1} z \,,\,\mbox{for} \, M^2=-\frac{3}{L^2}\,,\\
\mbox{v} (z)&=& \frac{\mbox{v}_1}{L_1^2} z^2 \,,\,\mbox{for} \, M^2=-\frac{4}{L^2}\,.
\eeqs
At this point,  I take the limits (after proper regularization) $L_0\rightarrow L \rightarrow 0$.

As a result, the two cases I discuss correspond to 
the assumption of having a chiral condensate with dimensions
$d=1$ and $d=2$ respectively. The first is somewhat equivalent
(from the electro-weak scale EFT point of view)
to a four-dimensional model in which symmetry breaking
is triggered by a physical scalar Higgs (composite), the 
second to the case of walking TC, in which symmetry breaking is induced by 
a $\langle\bar{\psi}\psi\rangle$ techni-quark condensate with large anomalous dimension
($d=2$). Both are relevant phenomenological choices
in the dynamical electro-weak symmetry breaking context.

Notice how the choice of setting $B=0$ does not violate
unitarity bounds in these two cases~\cite{BF}. 
The choice $M^2=-4/L^2$ saturates the lowest bound
on the possible mass for the bulk scalar~\cite{witten}. In this case,
the choice $B=0$ is the only one compatible with conformal symmetry.
For $M^2=-3/L^2$, in principle both choices of $B=0$ or $A=0$ 
give rise to consistent theories~\cite{KW}.  The $d=3$ case
has been studied at length in the literature,
being the natural choice in the $AdS/QCD$, because
 the QCD condensate $\langle \bar{q}q \rangle$ is 
 represented by a solution to the bulk equations with scaling dimension
 $d=3$.
 This has the disadvantage that 
 the bulk equations for the axial spin-1 fields  can be solved only numerically.
The choice $d=1$ is, instead, exactly solvable, but requires some 
attention in the regularization procedure, in order
to make the zero-modes normalizable
(and hence allowing for non-vanishing couplings with the tower
of excited states) without introducing tachyonic 
degrees of freedom.

%%%%%%%%%%%%%%%%%%%%%%%%%%%%%%%%%%%%%%%%%%%%%%%%%%%%%%%%%%%%%%%%%%%%%%
\section{Preliminaries.}
%%%%%%%%%%%%%%%%%%%%%%%%%%%%%%%%%%%%%%%%%%%%%%%%%%%%%%%%%%%%%%%%%%%%%%

Proper quantization of the gauge theory requires the introduction
of gauge-fixing terms:
\beqs
{\cal L}_{GF}&=& \frac{1}{\xi_V}\frac{L}{z}\Tr\left[\partial_{\mu}V^{\mu}-\frac{z}{L}\xi_{V}\partial_z \frac{L}{z}V_5\right]\,\nonumber\\
&&+ \frac{1}{\xi_A}\frac{L}{z}\Tr\left[\partial_{\mu}A^{\mu}-\frac{z}{L}\xi_{A}\partial_z \frac{L}{z}A_5-\xi_A\frac{L^2}{z^2}\frac{\sqrt{g^2+g^{\prime\,2}}}{2}\mbox{v}\phi_{0}\right]\,,\\
&&+ \frac{1}{\xi_+}\frac{L}{z}\Tr\left[\partial_{\mu}W^{\mu}-\frac{z}{L}\xi_{+}\partial_z \frac{L}{z}W_5-\xi_+\frac{L^2}{z^2}\frac{g}{2}\mbox{v}\phi_{W}\right]\,.\nonumber
\eeqs
 Here and in the following,
$V=V_3T_3$, $A=A_3T_3$ and $W=W_1T_1+W_2T_2$
are defined in terms of the corresponding components of the original fields as:
\beqs
V &=&\frac{g^{\prime} L\,+\,g R}{\sqrt{g^2+g^{\prime 2}}}\,,\\
A &=&\frac{g L\,-\,g^{\prime} R}{\sqrt{g^2+g^{\prime 2}}}\,,\\
W &=& L\,.
\eeqs
Analogously,
$\phi_0\equiv \phi_3T_3$ and $\phi_{W}\equiv \phi_1T_1+\phi_2T_2$.

These choices of gauge fixing allow to cancel the bilinear mixing between spin-1 and spin-0 
fields arising from the bulk action. 
With this, the action 
can be written as the sum over  vectorial part $S_V$ and
axial part $S_A$ (the $S_{+}$ part is identical to the $S_A$ 
after replacing $A\rightarrow W$, $g^2+g^{\prime\,2}\rightarrow g^2$,
$\xi_A\rightarrow \xi_{+}$
and $\phi_0\rightarrow \phi_W$):

\begin{widetext}
\beqs
{\cal S}_{V} &=& \int\di^4 x \int_{L_0}^{L_1}\di z\,\frac{L}{z}\left[\frac{}{}
1+D\,\delta(z-L_0)\right]\left[ -\frac{1}{2}\Tr V_{\mu\nu}V^{\mu\nu}\right]\nonumber\\
&& +\frac{L}{z}\Tr\,\partial_z V_{\mu} \partial_z V^{\mu}\,
-\,\frac{1}{\xi_V}\frac{L}{z}\Tr\left[\partial_{\mu}V^{\mu}\right]^2
\,\label{SV}\\
&&+\frac{L}{z}\Tr\left[\partial_{\mu}V_5\partial^{\mu}V_5\right]
\,-\,\xi_V\frac{z}{L}\Tr\left[\partial_{z}\frac{L}{z}V_5\right]^2\nonumber\\
&&+2\partial_z\left\{\frac{L}{z}\Tr\left[\partial_{\mu}V^{\mu}V_5\right]\right\}\nonumber\,,\\
{\cal S}_{A} &=& \int\di^4 x \int_{L_0}^{L_1}\di z\,\frac{L}{z}\left[\frac{}{}
1+D\,\delta(z-L_0)\right]\left[ -\frac{1}{2}\Tr A_{\mu\nu}A^{\mu\nu}\right]\nonumber\\
&& +\frac{L}{z}\Tr\,\partial_z A_{\mu} \partial_z A^{\mu}\,
-\,\frac{1}{\xi_A}\frac{L}{z}\Tr\left[\partial_{\mu}A^{\mu}\right]^2
\,\label{SA}\\
&&+\,\left(\frac{L}{z}\right)^3\left[1+C\delta(z-L_0)\right]\frac{g^2+g^{\prime\,2}}{4}\mbox{v}^2\Tr\left[A_{\mu}A^{\mu}\right]\,\nonumber\\
&&+\frac{L}{z}\Tr\left[\partial_{\mu}A_5\partial^{\mu}A_5\right]\,\,\nonumber\\
&&-\,\xi_A\frac{z}{L}\Tr\left[\partial_{z}\frac{L}{z}A_5+\frac{\sqrt{g^2+g^{\prime\,2}}}{2}\left(\frac{L}{z}
\right)^3\mbox{v}\phi_0\right]^2\nonumber\\
&&+2\partial_z\left\{\frac{L}{z}\Tr\left[\partial_{\mu}A^{\mu}A_5\right]\right\}\,
+\,\left(\frac{L}{z}\right)^3C\delta(z-L_0)\mbox{v}\sqrt{g^2+g^{\prime\,2}}\Tr\left[A^{\mu}\partial_{\mu}\phi_0\right]\nonumber\,\\
&&+\left(\frac{L}{z}\right)^3\left[1+C\delta(z-L_0)\right]\Tr\left[\partial_{\mu}\phi_0\partial^{\mu}\phi_0\right]\nonumber\\
&&-\,\left(\frac{L}{z}\right)^3\frac{\mbox{v}^2}{g^2+g^{\prime\,2}}
\Tr\left[\frac{g^2+g^{\prime\,2}}{2}A_5
+\sqrt{g^2+g^{\prime\,2}}\partial_z\frac{\phi_0}{\mbox{v}}\right]^2\nonumber
\eeqs
\end{widetext}
The unitary gauge is defined by $\xi_i=+\infty$. 

The wave equations deduced from ${\cal S}_V$ for the pseudo-scalar 
in unitary gauge, together with the boundary terms, are satisfied by:
\beqs
V_5&=&0\,.
\eeqs
Hence one finds the action for the spin-1 states to reduce to
\beqs
{\cal S}_{V} &=& \int\di^4 x \int_{L_0}^{L_1}\di z\,\frac{L}{z}\left[\frac{}{}
1+D\,\delta(z-L_0)\right]\left[ -\frac{1}{2}\Tr V_{\mu\nu}V^{\mu\nu}\right]\nonumber\\
&& +\frac{L}{z}\Tr\,\partial_z V_{\mu} \partial_z V^{\mu}\,,\\
&=&  \int\di^4 q \int_{L_0}^{L_1}\di z\,\frac{L}{z}
\Tr V^{\mu}\left[\left(1+D\delta(z-L_0)\right)q^2 P_{\mu\nu}
\,+\,\eta_{\mu\nu}\frac{z}{L}\partial_z\frac{L}{z}\partial_z\right]V^{\nu}\nonumber\\
&&+\int\di^4 q \int_{L_0}^{L_1}\di z\,\partial_{z}\left(\frac{L}{z}\Tr V_{\mu}\partial_z V^{\mu}\right)\,.
\eeqs
After Fourier transforming in the four-dimensional coordinates,
one can factorize the dependence on the fifth coordinate, and write:
$V^{\mu}(q,z)=V^{\mu}(q)v_v(z,q)$. Imposing the bulk equations
\beqs
\frac{z}{L}\partial_z\frac{L}{z}\partial_z v_v(z,q)&=&-q^2 v_v(z,q)\,,
\eeqs
and using Neumann bounday conditions in the IR
\beqs
\partial_z v_v(L_1,q) &=& 0\,,
\eeqs
the action reduces to the action at the UV boundary:
\beqs
{\cal S}_{\partial V} &=&
  \int\di^4 q \Tr V^{\mu}\left[ \int\di z\,\frac{L}{z}
  \delta(z-L_0)v_v(z,q) 
  \left((D q^2 P_{\mu\nu} v_v(z,q)+ \eta_{\mu\nu}\partial_z v_v(z,q))\right)\right] V^{\nu}\,,
\eeqs
from which, introducing appropriate localized currents for the SM gauge fields,
the polarization  appears to be:
\beqs
\pi_v \,=\,\frac{1}{p_v} &=&{\cal N}\left(Dq^2+\frac{\partial_zv_v}{v_v}\right)(L_0,q)\,,
\eeqs
with the overall normalization 
\beqs
{\cal N}^{-1} &\equiv & \int_{L_0}^{L_1}\di z\,\frac{L}{z}\left[\frac{}{}
1+D\,\delta(z-L_0)\right]\,\nonumber\\
&=&L\left(\frac{D}{L_0}+\ln\frac{L_1}{L_0}\right)\,.
\eeqs

From ${\cal S}_A$, the unitary gauge implies:
\beqs
\phi_0&=&-\left(\frac{z}{L}\right)^3\frac{2}{\sqrt{g^2+g^{\prime\,2}}\mbox{v}}\partial_z\frac{L}{z}A_5\,,
\eeqs
and the boundary conditions for $A_5$ are derived from the 
condition of not having mixing at the boundaries:
\beqs
A_{5} (L_1)&=& 0\,,\\
\left(\frac{L}{z}A_5-C\partial_z\frac{L}{z}A_5\right)(L_0)&=&0\,.
\eeqs
The action for the spin-1 fields is hence, in unitary gauge:
\beqs
{\cal S}_{A} &=& \int\di^4 x \int_{L_0}^{L_1}\di z\,\frac{L}{z}\left[\frac{}{}
1+D\,\delta(z-L_0)\right]\left[ -\frac{1}{2}\Tr A_{\mu\nu}A^{\mu\nu}\right]\nonumber\\
&& +\frac{L}{z}\Tr\,\partial_z A_{\mu} \partial_z A^{\mu}\,\\
&&+\,\left(\frac{L}{z}\right)^3\left[1+C\delta(z-L_0)\right]\frac{g^2+g^{\prime\,2}}{4}\mbox{v}^2\Tr\left[A_{\mu}A^{\mu}\right]\,'\nonumber\\
\eeqs
very similar to the previous case, but for the presence of the symmetry breaking terms
induced by the non-vanishing $\mbox{v}(z)$.
Defining $A^{\mu}(q,z)\equiv A^{\mu}(p)v_a(z,q)$, and redoing the same 
procedure as above, the polarization is:
\beqs
\pi_a \,=\,\frac{1}{p_a} &=&{\cal N}\left(Dq^2+\frac{\partial_zv_a}{v_a}-C\left(\frac{L}{z}\right)^2\frac{g^2+g^{\prime\,2}}{4}\mbox{v}^2\right)(L_0,q)\,,
\eeqs
where $v_a$ satisfies
\beqs
\frac{z}{L}\partial_z\frac{L}{z}\partial_z v_a(z,q)-\left(\frac{L}{z}\right)^2\frac{g^2+g^{\prime\,2}}{4}\mbox{v}^2v_a(z,q)&=&-q^2 v_a(z,q)\,,\\
\partial_zv_a(L_1,q)&=&0\,,
\eeqs
and where the normalization constant  ${\cal N}$ is the same as above.
The analysis of the ${\cal S}_{+}$ is the same, with the due substitutions.

The boundary terms $C$ and $D$ have a very important role, that
is well illustrated by looking at the wave functions for the zero modes
of the vector bosons and of the pions.

Focusing of the ${\cal S}_A$ sector  in unitary gauge, 
and writing $A_5\equiv A_5(q) f(z,q)$,
the zero-modes satisfy the equation
\beqs
\partial_z\left(\frac{z}{L}\right)^3\frac{1}{\mbox{v}^2}\partial_z\frac{L}{z} f(z,0)&=&\frac{1}{4}(g^2+g^{\prime\,2})f(z,0)\,,
\eeqs
with the boundary conditions 
\beqs
f(L_1,0)&=&0\,,\\
\left(\frac{L}{z}f(z,0)-C\partial_z\frac{L}{z}f(z,0)\right)(z=L_0)&=&0\,,
\eeqs
and the normalization given by
\beqs
1&=&\int_{L_0}^{L_1}\di z\,\left[\frac{L}{z}
f(z,0)^2+\left(\frac{z}{L}\right)^3\frac{4(1+C\delta(z-L_0))}{(g^2+g^{\prime\,2})^2\mbox{v}^2}(\partial_z f(z,0))^2\right]\,.
\eeqs

In  the two cases discussed here, it is convenient to write
\beqs
\mbox{v}(z)&=& a z^d\,,
\eeqs
 and then
rewrite $M_Z^2=(g^2+g^{\prime\,2})^2a^2L^2/4$.
The solutions, having imposed the boundary conditions in the IR, can be written as
\beqs
f(z,0)&=&c_0 z \left(K_0(M_ZL_1)I_0(M_Zz)-I_0(M_ZL_1)K_0(M_Zz)\right))\,,\,{\rm for}\,d=1\,,\\
f(z,0)&=&c_0 z e^{-M_Zz^2/2}\left(1-e^{M_Z(z^2-L_1^2)}\right)\,,\,{\rm for}\,d=2\,,
\eeqs
with the constant $c_0$ determined by the normalization conditions.

Taking the limit $L_0\rightarrow L\rightarrow 0$ is somewhat problematic,
and will be discussed explicitly in the two examples later on, showing 
that this procedure dictates the form of the counter-terms $C$ and $D$. 
From Eq.(\ref{SA}), one can see that the counter-term $C$ enters in the overall 
normalization of the pion state.  Unitarity requires such normalization to be positive and 
finiteness of the coupling to other states requires it to be finite,
and hence constraints the value of $C$. Similarly, $D$ enters the normalization
of the photon field, and hence finiteness of the gauge coupling requires a finite
normalization, while unitarity requires positive normalization.

It is  useful to consider
 the limit in which in the dual strongly interacting  theory the chiral symmetry is
 global, and the model becomes the effective description of a QCD-like theory
 (i.e., in which the weak gauging of the global symmetry of the CFT is set to zero).
In order to do this, remind that the propagator of the photon can be expressed in terms
of the generating functional for Green functions $\Sigma(q)$,
computed by treating the value of the gauge bosons at the UV-boundary
as a (non-dynamical) source coupled to the currents $J_{\mu}$:
\beqs
\langle J_{\mu} J_{\nu}\rangle & = &i\, P_{\mu\nu}\, \Sigma(q)\,,
\eeqs
 in the form~\cite{pheno}
\beqs
\pi_{v}(q^2) &=& q^2-g_4^2\,\Sigma(q)\,.
\eeqs
Hence, the current-current correlators of the QCD-like limit for these models
can be computed as
\beqs
\Sigma_{VV}(q^2)&=&\frac{1}{g_4^2}\left(q^2-\pi_v(q^2)\right)\,,\\
\Sigma_{AA}(q^2)&=&\frac{1}{g_4^2}\left(q^2-\pi_a(q^2)\right)\,,
\eeqs
with $g_4$ the effective
gauge coupling in four dimensions of the zero-modes.

These correlators can be written (up to divergent constants) as:
\beqs
\Sigma_{VV}(q^2)&=&q^2\sum_n\frac{f_{\rho\,n}^2}{q^2-M_{\rho\,n}^2}\,,\\
\Sigma_{AA}(q^2)&=&f_{\pi}^2+q^2\sum_n\frac{f_{a_1\,n}^2}{q^2-M_{a_1\,n}^2}\,,
\eeqs
where the first sum  runs over the vector mesons,
and the second over the axial-vector mesons. The poles 
of the correlators give the masses of $\rho$ mesons and $a_1$ mesons,
with all their excited states, and the residues the decay constants, while $f_{\pi}^2$
is the pion decay constant.

%%%%%%%%%%%%%%%%%%%%%%%%%%%%%%%%%%%%%%%%%%%%%%%%%%%%%%%%%%%%%%%%%%%%%%
\section{$AdS_5$ Background, $d=1$}
%%%%%%%%%%%%%%%%%%%%%%%%%%%%%%%%%%%%%%%%%%%%%%%%%%%%%%%%%%%%%%%%%%%%%%
 I consider first the case with $d=1$, i.e.:
\beqs
\mbox{v}(z)&=&\frac{\mbox{v}_1}{L_1}z\,.
\eeqs

The solution of the bulk equations, after imposing the Neumann 
boundary conditions in the IR, reads:
\beqs
v_i(z,q)&=&c_i z \left(J_0(k_iL_1)Y_1(k_iz)-Y_0(k_iL_1)J_1(k_iz)\right)\,,
\eeqs
where $i=v,a,+$, $c_i$ are normalization constants, and where
$k_v=q$, $k_{+}=\sqrt{q^2-M_W^2}$ and $k_a=\sqrt{q^2-M_Z^2}$.

Consider first the vectorial sector. The polarization tensor,
using the formulae discussed in the previous section, is given by:
\beqs
\pi_v(q^2)&=&\left.\frac{Dq^2+\partial_zv_v(z,q)/v_v(z,q)}{L(D/L+\ln L_1/L_0)}\right|_{z=L_0}\,.
\eeqs
Taking $L_0\rightarrow L$, defining $D\equiv L D^{\prime}$, and expanding 
for $L\rightarrow 0$:
\beqs
\pi_v(q^2)&=&\frac{q^2}{D^{\prime}+\ln \frac{L_1}{L}}\left(\frac{\pi}{2}
\frac{Y_0(qL_1)}{J_0(qL_1)}-\left(\gamma_E+\ln \frac{qL}{2} -D^{\prime}\right)\right)\,,
\eeqs
which contains divergent terms for $L\rightarrow 0$. These can be reabsorbed in the
localized boundary terms, by defining:
\beqs
\label{regulator}
D^{\prime}&\equiv&\ln \frac{L}{L_1} + \frac{1}{\varepsilon^2}\,,
\eeqs
and hence concluding that
\beqs
\pi_v(q^2)&=&q^2\left(1-\varepsilon^2\left(
\gamma_E+\ln \frac{qL_1}{2}-
\frac{\pi}{2}
\frac{Y_0(qL_1)}{J_0(qL_1)}\right)\right)\,.
\eeqs

The meaning of this procedure can be understood by
looking at the explicit expression of the wave function
of the  zero mode of the vectorial part (to be identified with the photon),
which with these definitions  is given by
\beqs
v_v^{(0)} & = & \frac{\varepsilon}{\sqrt{L}}\,.
\eeqs
By contrast, for a constant $v_v^{(0)}(z)=C^{(0)}$:
\beqs
\int_{L_0}^{L_1}\di z \, \frac{L}{z} \,C^{(0)\,2} &=& L\log\frac{L_1}{L_0}  \,C^{(0)\,2}\,.
\eeqs
Without the introduction of $D$, the normalization condition would introduce a 
spurious logarithmic dependence on the unphysical UV cut-off scale $L_0$.
It would not be possible to take the limit $L_0\rightarrow L \rightarrow 0$ without 
effectively decoupling the photon from the spectrum (this been a {\it non normalizable} mode).
To retain non-vanishing couplings of the photon, it would be necessary to work with finite $L_0$.
This would affect all the physical quantities with an explicit, divergent, dependence on
the precise choice of the cut-off (and of the regularization procedure itself,
since there is no reason to think that a hard-wall cut-off in the UV has any physical meaning). 
The presence of the first term in Eq.~(\ref{regulator}) cancels this spurious dependence, trading this with a physical quantity $\varepsilon$, 
which encodes the difference
of the couplings of the zero modes and of the heavy modes to the brane.

The (five-dimensional) gauge coupling vanishes as $g \sim \sqrt{L}$ for
$L\rightarrow 0$.
The four-dimensional, standard-model weak couplings $g_4$ and $g_4^{\prime}$
can be read off the tree-level cubic and quartic
 interactions among the zero-modes and are given by
\beqs
g_4 &=& \frac{g \varepsilon}{\sqrt{L}}\,,\\
g^{\prime}_4 &=& \frac{g^{\prime} \varepsilon}{\sqrt{L}}\,,
\eeqs
so that, at fixed $\varepsilon$, they are finite for $L\rightarrow 0$.
This means that the procedure adopted here is equivalent to
taking the limit in which the UV cut-off goes to infinity 
by keeping the electro-weak gauge couplings fixed.

The procedure for the other sectors is similar,\beqs
\pi_a(q^2)&=&
\left.\frac{Dq^2-CM_a^2+\partial_zv_a(z,q)/v_a(z,q)}{L(D/L+\ln L_1/L_0)}\right|_{z=L_0}\,,
\eeqs
 but for the fact that
new divergences arise, in connection with the symmetry breaking term,
that can be reabsorbed in $C$,
\beqs
\label{regulator2}
C&\equiv&L\left(\ln \frac{L}{L_1}+\frac{1}{\rho^2}\right)\,,
\eeqs
yielding
\begin{widetext}
\beqs
\pi_a(q^2)&=&q^2-\varepsilon^2\left(\frac{M_a^2}{\rho^2}+(q^2-M_a^2)\left(
\gamma_E+\ln\frac{L_1\sqrt{q^2-M_a^2}}{2}-\frac{\pi}{2}\frac{Y_0(L_1\sqrt{q^2-M_a^2})}{J_0(L_1\sqrt{q^2-M_a^2})}\right)\right)\,.
\eeqs
\end{widetext}
The expressions for the charged sector are identical, up to the replacement $M_a\rightarrow M_{+}$. 

The spectrum can be read off  the poles of the propagators.
Assuming the existence of a hierarchy between confinement scale
and masses of the gauge bosons,
i.e. $M_iL_1\ll1$:
\beqs
M_{W,Z}^2&\simeq&\frac{\varepsilon^2}{\rho^2}M_{+,a}^2\,.
\eeqs
The mass of the neutral techni-$\rho$'s is controlled by the confinement scale $L_1$ and 
depends on $\varepsilon$. The mass of the lightest such state can be approximated by:
\beqs
\left\{\begin{array}{cc}
M_{\rho^0}\,\simeq\,\frac{2.4}{L_1} \,,&\mbox{if}\, \varepsilon \ll  1\cr\cr
M_{\rho^0}\,\simeq\,\frac{4.1}{L_1} \,,&\mbox{if}\,  \varepsilon =  1\cr\cr
M_{\rho^0}\,\simeq\,\frac{4.7}{L_1} \,,&\mbox{if}\,  \varepsilon \gg 1\,.
\end{array}\right.
\eeqs

The precision observables are computed starting from 
the series expansion of the polarization tensors in the momenta:
\beqs
\pi_v(q^2)&\simeq& q^2\,+\,{\cal O}(q^4)\,,\\
\pi_a(0)&=&\varepsilon^2M_a^2\left(\gamma_E+\ln\frac{M_aL_1}{2}-\frac{1}{\rho^2}+\frac{K_0(M_aL_1)}{I_0(M_aL_1)}\right)\,,\\
\pi_a^{\prime}(0)&=&1-\frac{\varepsilon^2}{2}\left(1+2\gamma_E+\ln\frac{L_1^2M_a^2}{4}
-\frac{1-2K_0(M_aL_1)I_0(M_aL_1)}{I_0(M_aL_1)^2}\right)\,\\
&=&1-\frac{\varepsilon^2}{2}\left(1+\frac{2}{\rho^2}-\frac{1}{I_0(M_aL_1)^2}\right)-\frac{\pi_a(0)}{M_a^2}
\eeqs

\begin{widetext}
\beqs
\hat{S}&=&
\frac{ \varepsilon^2  M_W^2}{2  M_Z^2} \left(\frac{2
   I_0\left( L_1 M_a\right)
   K_0\left( L_1
    M_a\right)-1}{I_0\left( L_1
    M_a\right)^2}+\log
   \left(\frac{ L_1^2  M_a^2}{4}\right)+2 \gamma_E
   +1\right)
   \\
   &\simeq&\frac{1}{2} \varepsilon^2  \frac{M_W^2}{M_Z^2}M_a^2L_1^2
   \,=\,\frac{\rho^2}{2}M_W^2L_1^2\,,\\
\hat{T}
   &\simeq&\frac{\rho^4}{4\varepsilon^2}L_1^2\left(M_Z^2-M_W^2\right)\,,
\eeqs
\end{widetext}
where the approximations are valid for $M_{W,Z}\ll1/L_1$.
The ratio of the two is independent of the confinement scale:
\beqs
\frac{\hat{T}}{\hat{S}}&\simeq &\frac{M_Z^2-M_W^2}{2M_W^2}\frac{\rho^2}{\varepsilon^2}\,\simeq\,0.15\frac{\rho^2}{\varepsilon^2}\,.
\eeqs

In order to have a better understanding of what the boundary terms mean,
and also in order to illustrate why the spectrum depends on $\varepsilon$,
I turn the attention to the limit in which the strongly and weakly coupled sectors decouple
from each other. The spectrum of  the strongly interacting sector of the model reduces 
to a QCD-like spectrum, containing a set of pions and two towers of heavy $\rho$ and $a_1$
states, decoupled from the electro-weak gauge bosons (photon, $W$ and $Z$ are non normalizable in this limit). This is obtained in the limit $\varepsilon\rightarrow 0$.

The vector-vector correlator is:
\beqs
\Sigma_{VV}(q^2)&=&\frac{1}{g_4^2}\left(q^2-\pi_v(q^2)\right)\,\\
&=&\frac{1}{g_{\rho}^2} q^2\left(\gamma_E+\ln\frac{qL_1}{2}-\frac{\pi}{2}\frac{Y_0(qL_1)}
{J_0(qL_1)}\right)\,\\
&=&q^2\left(\sum_n\frac{q^2f_{\rho\,n}^2}{M_{\rho\,n}^2(q^2-M_{\rho\,n}^2)}\right)\,,\\
&=&q^2\left(\sum_n\frac{f_{\rho\,n}^2}{M_{\rho\,n}^2}+\sum_n\frac{f_{\rho\,n}^2}{q^2-M_{\rho\,n}^2}\right)\,,
\eeqs
from which one derives the spectrum of the strong sector. The 
coupling $g_{\rho}=g/\sqrt{L}$ is the effective coupling of the vector mesons
in the four-dimensional language.
The masses of the vector mesons are given by the zeros of $J_0(qL_1)$
\beqs
M_{\rho\,n}&=&\frac{1}{L_1}\left(\frac{}{}2.4\,,\,5.5\,,\,8.6\,,\,11.8\,,\,\cdots\right)\,,\\
&\simeq&\frac{1}{L_1}(\pi(n-1/4))\,
\eeqs
and the residues of $\Sigma_{VV}$ give the decay constants:
\beqs
f_{\rho\,n}&=&\frac{1}{g_{\rho}L_1}\left(\frac{}{}2.7\,,\,4.1\,,\,5.2\,,\,6.1\,,\,\cdots\right)\,.\\
f_{\rho\,n}^2&\simeq&\frac{1}{g_{\rho}^2L_1^2}(-2.45+\pi^2n)\,.
\eeqs
Notice how a parametric suppression of the decay constants 
could be achieved by tuning by hand the coupling $g_{\rho}$ to very large values.

In the limit in which $q\gg1/L_1$, the correlator behaves as:
\beqs
\Sigma_{VV}(q^2)&\rightarrow&\frac{1}{2g_{\rho}^2}q^2\ln q^2\,,
\eeqs
which, compared with the OPE results, suggests that
\beqs
\frac{1}{g_{\rho}^2}&\simeq&\frac{N_c}{12\pi^2}\,,
\eeqs
with $N_c$ the number of colors of the $SU(N_c)$ QCD-like dynamics.
In all what done here, the basic assumption is that the five-dimensional
gauge coupling $g$ be small. This expression shows that this corresponds to
the limit of large $N_c$ of the dual  description, and hence indicates that
the computations performed here are accurate only in this regime.
In particular, this means that one cannot take $g_{\rho}$ to arbitrarily large
values, because this would invalidate the five-dimensional perturbative expansion 
used here, or equivalently, in the dual description, this would be equivalent to
the study of a small $N_c$ strongly coupled model, for which no
parametric suppression of the loop effects is present.

For the axial-axial correlator:
\beqs
\Sigma_{AA}(q^2)&=&\frac{1}{g_{\rho}^2}\left(\frac{M^2}{\rho^2}+(q^2-M^2)\left(\gamma_E+\ln\frac{L_1\sqrt{q^2-M^2}}{2}
-\frac{\pi}{2}\frac{Y_0(L_1\sqrt{q^2-M^2})}
{J_0(L_1\sqrt{q^2-M^2})}\right)\right)\,\\
&=&\frac{M^2}{g_{\rho}^2\rho^2}+\Sigma_{VV}(q^2-M^2)\,\\
&=&\frac{M^2}{g_{\rho}^2\rho^2}+(q^2-M^2)\left(\sum_n\frac{(q^2-M^2)f_{\rho\,n}^2}{M_{\rho\,n}^2(q^2-M^2-M_{\rho\,n}^2)}\right)\,\\
&=&\frac{M^2}{g_{\rho}^2\rho^2}-M^2\sum_n\frac{M^2f_{\rho\,n}^2}{M_{\rho\,n}^2(M^2+M_{\rho\,n}^2)}+q^2\sum_n\frac{f_{\rho\,n}^2}{M_{\rho\,n}^2}\\
&&+\,q^2\sum_n\frac{f_{\rho\,n}^2M_{\rho\,n}^2}{(M^2+M_{\rho\,n}^2)(q^2-M^2-M_{\rho\,n}^2)}\,,\nonumber
\eeqs
where $M^2=g^2L_1^2\mbox{v}_0^2/2$ (notice that in looking at the strong sector by itself,
it is convenient to look at the $g=g^{\prime}$ limit, directly comparable to QCD-like theories).
From here, the masses and decay constants of the 
axial excitations can be derived to be:
\beqs
M_{a_1\,n}^2&=&M_{\rho\,n}^2+M^2\,,\\
f_{a_1\,n}^2&=&f_{\rho\,n}^2\frac{M_{\rho\,n}^2}{M^2+M_{\rho\,n}^2}\,,
\eeqs
which automatically implies that one of the Weinberg sum rules 
is satisfied:
\beqs
\sum_n(f_{\rho\,n}^2M_{\rho\,n}^2-f_{a_1\,n}^2M_{a_1\,n}^2)&=&0\,.
\eeqs
The pion decay constant is~\footnote{This result comes from the regularization on the UV-brane. Removing completely the boundary term $C$, and taking the $L_0\rightarrow 0$ limit naively, the normalization of the pion field diverges. Analogously, if $D=0$
the normalizations of the photon, $W$ and $Z$ gauge bosons diverge.
This theory would consist of a set of massless, free 
gauge bosons and pseudo-scalars, with vanishing couplings to the 
SM currents. The regularization and renormalization procedure adopted here
requires a choice of $C$ and $D$ such as to make these normalizations both
finite and positive, so as to restore finite interactions while preserving unitarity. 
As such, it should not come as a surprise that both the
gauge coupling of the photon as well as the pion decay constant
are free parameters in the $d=1$ case.}

\beqs
f_{\pi}^2&=&\Sigma_{AA}(0)\,\\\label{pion}
&=&\frac{M^2}{g_{\rho}^2\rho^2}-M^4\sum_n\frac{f_{\rho\,n}^2}{M_{\rho\,n}^2M_{a_1\,n}^2}\,\\
&=&\frac{M^2}{g_{\rho}^2\rho^2}-M^2\left(\gamma_E+\ln\frac{ML_1}{2}+\frac{K_0(ML_1)}{I_0(ML_1)}\right)\,,
\eeqs
which allows to replace $\rho$ with the decay constant:
\beqs
\frac{1}{\rho^2}&=&g_{\rho}^2\left(\frac{f_{\pi}^2}{M^2}+\gamma_E+\ln\frac{ML_1}{2}+\frac{K_0(ML_1)}{I_0(ML_1)}\right)\,.
\eeqs
The other Weinberg sum rule is not satisfied:
\beqs
\sum_n\left(f_{\rho\,n}^2-f_{a1\,n}^2\right)-f_{\pi}^2&=&\sum_nf_{\rho\,n}^2\left(1-\frac{M_{\rho\,n}^2}{M_{\rho\,n}^2+M^2}-\frac{M^4}{M_{\rho\,n}^2(M_{\rho\,n}^2+M^2)}\right)-\frac{M^2}{g_{\rho}^2\rho^2}\,,\\
&=&M^2\left(-\frac{1}{g_{\rho}^2\rho^2}+\sum_n f_{\rho\,n}^2\frac{M_{\rho\,n}^2-M^2}{M_{\rho\,n}^2(M_{\rho\,n}^2+M^2)}\right)\,\\
&\neq&0\,,
\eeqs
unless $M=0$.
This is a test illustrating the fact that the interpretation of the 
setting given is the one anticipated, namely that
the condensate responsible for symmetry breaking 
has dimension $d=1$.

Going back to the precision observables, it is possible to trade
now the unknown parameter $\rho$ for $f_{\pi}$, and 
explicitly verify that
\beqs
\label{sSdispersion1}
\hat{S}&=&\varepsilon^2\frac{M_W^2}{f_{\pi}^2}\sum\left(\frac{f_{\rho\,n}^2}{M_{\rho\,n}^2}-
\frac{f_{\rho\,n}^2}{M_{\rho\,n}^2}\right)\,,
\eeqs
which relates the precision observables to the masses and decay constants of the
strong sector of the theory. In Eq.~(\ref{sSdispersion1}), masses and decay constants
are  not those of  the techni-mesons,
but the position of the poles in $\Sigma_{VV}$, giving the 
spectrum of the strong sector alone, decoupled from the SM
gauge bosons.
The quantity $\hat{S}$ is positive definite, as a result of the fact that
$\varepsilon^2$ has to be chosen to be positive in order 
to have a positive normalization of the photon wave function,
and hence  positivity of the spin-1 boson contribution
to $\hat{S}$ is a consequence of unitarity requirements.

Looking at the expression for $f_{\pi}$, or equivalently to 
$\pi_a(0)$, one sees that setting $1/\rho^2\rightarrow 0$
results in the model being pathological, since Eq.~(\ref{pion})
is not positive in this case. Nevertheless, the fact that 
$C$ is needed as a divergent counter-term also
implies that $1/\rho^2$ is a parameter in the model,
that has to be chosen as to make these quantities positive.
Setting $1/\rho\rightarrow 0$ would be  just a wrong choice for the regulator.
The model with $d=1$ has, hence, two  free parameters more
than the naive counting of the bulk interaction terms would suggest.
One controls the relative strength of the effective coupling of the 
SM gauge bosons in respect to the one of the excited states ($\varepsilon$).
The other controls the relative size of the symmetry breaking effects
experienced by the lowest modes in respect to those experienced by 
the excited states ($\rho$).

%%%%%%%%%%%%%%%%%%%%%%%%%%%%%%%%%%%%%%%%%%%%%%%%%%%%%%%%%%%%%%%%%%%%%%
\section{$AdS_5$ Background, $d=2$.}
%%%%%%%%%%%%%%%%%%%%%%%%%%%%%%%%%%%%%%%%%%%%%%%%%%%%%%%%%%%%%%%%%%%%%%

If the scalar background is
\beqs
\mbox{v}&=&\frac{\mbox{v}_1}{L_1^2}z^2\,=\,\frac{\mbox{v}_0}{L_0^2}z^2\,,
\eeqs
this has the effect of modifying the equations of motion in
the AdS background for the tower of excitations of $W$ and $Z$ gauge bosons:
\beqs
\partial_z\frac{L}{z}\partial_z v_v &=&-q^{2}\frac{L}{z}v_v\,,\\
\partial_z\frac{L}{z}\partial_z v_a-\mu^4_{Z} L z  v_a&=&-q^{2}\frac{L}{z}v_a\,,\\
\partial_z\frac{L}{z}\partial_z v_{+}-\mu^4_{W} L z  v_{+} &=&-q^{2}\frac{L}{z}v_{+}\,,
\eeqs
where  $\mu^4_W=1/4g^2\mbox{v}_0^2/L^2$ and $\mu^4_Z=1/4(g^2+g^{\prime 2})\mbox{v}_0^2/L^2$.  Notice that the term responsible for symmetry-breaking
has now a different $z$-dependence with respect to the mass term.

The general solution to the differential equation
can be written in terms of generalized Laguerre and Hypergeometric functions:
\beqs
v_a(z,q)&=&
e^{\frac{-\mu_Z^2z^2}{2}}\left[c_1U\left(-\frac{q^2}{4\mu_Z},0,\mu_Z^2 z^2\right)\,+\,
c_2L\left(\frac{q^2}{4\mu_Z^2},-1,\mu_Z^2z^2\right)\right]\,,
\eeqs
with $c_1$ and $c_2$ integration constants. Their ratio is fixed by
the IR-boundary conditions, and the overall normalization 
by the normalization of the states.
I choose to write them as:
\beqs
c_1&=&2L\left(-1+\frac{q^2}{4\mu_Z^2},\mu_Z^2L_1^2\right)+L\left(\frac{q^2}{4\mu_Z^2},-1,\mu_Z^2L_1^2\right)\,,\\
c_2&=&-U\left(-\frac{q^2}{4\mu_Z^2},0,\mu_Z^2L_1^2\right)+\frac{q^2}{2\mu_Z^2}
U\left(1-\frac{q^2}{4\mu_Z^2},1,\mu_Z^2L_1^2\right)\,.
\eeqs

On the UV boundary, for $L_0\rightarrow 0$:
\beqs
\frac{\partial_z a}{a}&\rightarrow&
L_0\left\{
\mu_Z^2\,-\,q^2\left[
\gamma_E+\ln(\mu_Z L_0)+\frac{1}{2}\psi\left(-\frac{q^2}{4\mu_Z^2}\right)-\frac{c_2}{2c_1}\Gamma\left(-\frac{q^2}{4\mu_Z^2}\right)
\right]\right\}\,,
\eeqs
with $\psi$ the digamma function.

The counter-term $D$ has been fixed by the normalization of the photon,
and hence:
\beqs
D&=&L_0\left(\ln\frac{L_0}{L_1}+\frac{1}{\varepsilon^2}\right)\,,\\
{\cal N}&=&\frac{\varepsilon^2}{L}\,.
\eeqs
Substituting this in the expression for the polarization, with the 
redefinition $C\equiv y^2/L_0$:
\beqs
\pi_a(q^2)&=&{\cal N}\left(Dq^2-C\mu_Z^4L_0^2+\frac{\partial_za}{a}(q^2,L_0)\right)\,\\
&=&
\frac{L_0}{L}\left[q^2-\varepsilon^2\left[-\mu_Z^2+y^2\mu_Z^4\frac{}{}\frac{}{}\right.\right.\\
&&\left.\left.+\, q^2\left[\ln\mu_Z L_1 +\gamma_E+\frac{1}{2}\psi\left(-\frac{q^2}{4\mu_Z^2}\right)
-\frac{c_2}{2c_1}\Gamma\left(-\frac{q^2}{4\mu_Z^2}\right)\right]\right]\right]\,,\nonumber
\eeqs
which shows how the regulator needed for the vector fields 
also regularizes the axial fields in this case.
Notice that, as expected, in the $\varepsilon\rightarrow 0$ limit
the polarization reduces to that of a single, massless spin-1 field.
Finally, it is now possible to take the limit $L_0\rightarrow L \rightarrow 0$,
which has been regularized, al long as $\mu_Z$ and  $y$ are kept fixed.

It is easier to discuss the properties of the
strong interacting sector of this model by  looking 
at $\Sigma_{AA}(q^2)$:
\beqs
\Sigma_{AA}(q^2)&=&
\frac{1}{g_{\rho}^2}\left[-\mu_Z^2+y^2\mu_Z^4+q^2\left[\gamma_E+\ln \mu_Z L_1+
\psi\left(-\frac{q^2}{4\mu_Z^2}\right)-\frac{c_2}{c_1}\Gamma\left(-\frac{q^2}{4\mu_Z^2}\right)
\right]\right]\,.
\eeqs
Taking the zero-momentum limit:
\beqs
f_{\pi}^2&=&\Sigma_{AA}(0)\,=\,\frac{\mu_Z^2}{g_{\rho}^2}\left(y^2\mu_Z^2
+\tanh\frac{\mu_Z^2L_1^2}{2}\right)\,.
\eeqs
Notice how $f_{\pi}^2$  is a positive-definite quantity in absence
of the counter-term $C$ ($y=0$), which 
proves the model being automatically unitary, in 
contrast with what found in the $d=1$ case.

The function $\Sigma_{AA}(q^2)$ can be plotted and studied easily,
in spite of its non-inspiring analytical expression. 
The explicit computation of residues and poles is
non trivial, though, and can be done only numerically.
It is hence worth discussing some special  limits, in which
the expressions simplify.
 
In the limit in which $\mu_Z^2L_1\rightarrow 0$, the bulk equations
reduce to those of the vectorial sector. Hence, for very small 
values of $\mu_Z^2L_1$, $\Sigma_{AA}\sim \Sigma_{VV}$.
In particular, the spectrum of the axial sector is given by the zeros
of the Bessel  function $J_0(qL_1)$, up to small corrections due to 
symmetry breaking. The position of the poles 
can be approximated by a quadratic sequence $M^2_{a_1\,n}\propto n^2$. 
The main deviation from $\Sigma_{VV}(q^2)$ is in the 
region $q^2\ll \mu_z^4L_1^2$, where
the bulk equations admit a simple solution in terms of hyperbolic functions,
which is the origin of the simple expression for $f_{\pi}$.

In the opposite limit, in which $\mu_z L_1^2 \gg 1$, the spectrum is severely
modified. Below $\mu_Z$,  the position of the poles grows linearly
$M^2_{a_1\,n} \propto n$, and then goes back to the quadratic 
behavior for very large masses $M_{a_1}\gg \mu_Z$.
To understand why, one can look explicitly at the expression for
the ratio $c_2/c_1$: in the limit of small $q^2$, and for asymptotically large
values of $\mu_Z^2L_1^2$, this vanishes, so that the correlator
reduces to
\beqs
\Sigma_{AA} &\sim& \frac{1}{g_{\rho}^2}\left(-\mu_Z^2+y^2\mu_Z^4
+q^2\left(\gamma_E+\ln{\mu_Z L_1}+\psi\left(-\frac{q^2}{4\mu_Z^2}\right)\right)\right)\,.
\eeqs
For small momenta, the poles are those of the $\psi(-q^2/(4\mu^2_Z))$,
i.e. the poles of the $\Gamma$ function. 
This property emerged in a similar model discussed in~\cite{karch},
and  was used there to reproduces the Regge trajectories.
Incidentally, the Veneziano amplitude reproduces 
the Regge trajectories for the same reason, its non-analytic
structure being dictated by the Euler Gamma function.
Of course, here this property applies only to the
axial sector, and hence this interpretation is not viable.

The actual computation of the decay constants is not
simple in this model, and not very illuminating.
It is hence difficult to verify 
the Weinberg sum rules. However, as long as  the AdS-CFT interpretation holds,
one expects that the second such rule be violated,
due to the presence of a dimension-2 condensate.  
This is analogous to what was studied in~\cite{AS}
in  the model-building effort to reduce $\hat{S}$ using  
the dispersion relations in order to quantify the non-perturbative 
corrections to the perturbative estimates.
As for the first sum rule, if the counter-term $C$ is assumed to scale
as $C \sim L_0$ (which is the natural choice), 
it disappears from the polarizations, and hence the
only symmetry breaking term would be the bulk VEV. In this 
way the first Weinberg sum rule  must hold.
Not so if one assumes the scaling $C = y^2/L_0$, with 
$y^2$ kept fixed when taking the limit $L_0\rightarrow L \rightarrow 0$.
This would imply the presence of an additional, dimension-1,
symmetry breaking term localized on the UV-brane,
possibly to be interpreted as the VEV of an additional
Higgs field, localized on the brane and not connected to the
strong sector. This might be a useful tool for the construction
of realistic mass matrices for the SM fermions.  But
the phenomenology would be determined by the interplay
between these two terms, one of which is not 
dictated by any specific  reason,  since $C$ is here neither
required by the regularization procedure
nor by unitarity arguments, and hence
from now on I will set $y=0=C$ in the $d=2$ case.

For the actual computation of the precision observables, 
it is useful to expand the polarization functions in powers of the momentum, obtaining:
\beqs
\pi_a(0)&=&-\varepsilon^2\mu_Z^2\left(\tanh\frac{\mu_Z^2L_1^2}{2}\right)\,.
\eeqs

In the limit of large $\mu^4L_1^4$, the $q^2$ coefficient of the expansion
can be approximated as:
\beqs
\pi^{\prime}_a(0) &\sim&\left(1-\frac{\varepsilon^2}{2}\left(\gamma_E+\ln\mu_Z^2 L_1^2\right)\right)\,,
\eeqs
while in the more interesting regime  in which $\mu_ZL_1\ll1$,
it is approximated by:
\beqs
\pi_a^{\prime}(0)&\simeq&1-\frac{\varepsilon^2}{2e}\mu_Z^4L_1^4\,,
\eeqs
with $e\simeq 2.7$.

If $\mu_Z L_1$ is large, then one obtains for $\hat{S}$:
\beqs
\label{largeS}
\hat{S}&=&\frac{\varepsilon^2}{2}\frac{M_W^2}{M_Z^2}\left(\gamma_E+\ln\mu_Z^2L_1^2\right)\,,
\eeqs
which is an ${\cal O}(1)$ number, and hence totally incompatible with
the experimental constraints. This case is hence excluded by experimental data
on precision electro-weak observables.

In the limit in which the symmetry breaking term
is small ($\mu_ZL_1^2\ll 1$) and in which the coupling 
between weak and strong sector is small ($\varepsilon \ll 1$) the
spectrum consists of a tower of heavy spin-1 fields, in which 
the mass splitting between $Z$, $W$ and photon excited states
is negligibly small, and the lightest of which has a  mass:
\beqs
M_{\rho^0}&=&\frac{2.4}{L_1}\,,
\eeqs
while the only light states are the SM gauge bosons, whose masses
are given by:
\beqs
M_{\gamma}^2&=&0\,\\
M_Z^2&\simeq&\frac{\varepsilon^2}{2}\mu_Z^4L_1^2\,,\\
M_W^2&\simeq&\frac{g^2}{g^2+g^{\prime\,2}}M_Z^2\,.
\eeqs

Finally, the precision parameters are given by
\beqs
\hat{S}&\simeq&\frac{\varepsilon^2}{2e}\frac{M_W^2}{M_Z^2}\mu_Z^4L_1^4\,\\
&\simeq&\frac{1}{e}M_W^2L_1^2\,,\\
\hat{T}&=&\varepsilon^2
\left(-\frac{\mu_Z^2}{M_Z^2}\tanh\frac{\mu_Z^2L_1^2}{2}
+\frac{\mu_W^2}{M_W^2}\tanh\frac{\mu^2_WL_1^2}{2}\right)\,\\
&\simeq&\frac{M_Z^2-M_W^2}{6\varepsilon^2}L_1^2\,,
\eeqs
so that  the ratio
\beqs
\frac{\hat{T}}{\hat{S}}&\simeq&\frac{e}{3\varepsilon^2}\frac{M_Z^2-M_W^2}{2M_W^2}\,,
\eeqs
does not depend on the confinement scale, as in the previous case.

\

%%%%%%%%%%%%%%%%%%%%%%%%%%%%%%%%%%%%%%%%%%%%%%%%%%%%%%%%%%%%%%%%%%%%%%
\section{Electro-weak Precision Parameters and Spin-1 Excitations.}
%%%%%%%%%%%%%%%%%%%%%%%%%%%%%%%%%%%%%%%%%%%%%%%%%%%%%%%%%%%%%%%%%%%%%%

The precision parameter  $\hat{T}$ is a very model-dependent quantity,
and in particular can be set to zero by modifying  the model
so that the excited states do not violate
custodial symmetry, for by gauging a full $SU(2)_L\times SU(2)_R$ 
in the bulk, and adding some additional symmetry breaking term
on the UV-brane.
Furthermore, the experimental bounds on $\hat{T}$ are easy to satisfy, 
because less stringent than the bounds on $\hat{S}$,
and because in these models the ratio $\hat{T}/\hat{S}$ is always 
suppressed by the ratio $(M_Z^2-M_W^2)/2M_W^2\simeq 0.15$.
The only problematic regime would be the one in which $\varepsilon$ is
very small, in which case all the computations performed here
would not be reliable anyhow.
For these reasons, I focus the discussion here on $\hat{S}$.

For the case $d=1$, I derived the approximate expression
\beqs
\hat{S}&\simeq&\frac{\rho^2}{2} M_W^2 L_1^2\,,\\
&=&\frac{k^2\rho^2}{2}\frac{M_W^2}{M_{\rho}^2}\,,
\eeqs
where the constant $k\in[2.4,4.7]$ is determined by the value of $\varepsilon$.
The experimental bound on $\hat{S}$ implies that
\beqs
M_{\rho^0}&>&k \rho (1 {\rm TeV})\,.
\eeqs

For small values of $\varepsilon$ and $\rho$, this is obviously satisfied
even for very light masses  $M_{\rho} \sim 1$ TeV.
For all practical purposes, $\hat{S}$ in this model is a free parameter,
only constrained to be positive by unitarity requirements.
The mechanism that makes $\hat{S}$ small by choosing small values
for $\rho$ essentially corresponds to a localization of the symmetry breaking
effects to the UV-brane, so that the SM gauge bosons are directly affected
by it, while the excited modes experience symmetry breaking 
with an additional suppression factor.

The spectrum of the model contains,
besides the usual SM gauge fields, with SM-like phenomenology,
a tower of spin-1 states, with tiny splitting between the excitations of 
the photon, the $W$ and the $Z$  bosons. The lightest 
such states have masses in the LHC energy range. 

The production (and decay) rates at the LHC are well illustrated
by looking at  the $g_{\rho\pi\pi}$ coupling of the strong sector.
In models in which the tower of excited spin-1 states 
is described by a (local) extra-dimension theory,
the KSRF relation is modified to~\footnote{A precise computation
of the value of $g_{\rho\pi\pi}$ requires to explicitly compute the 
3-point functions, which goes beyond the aims of this work.}
\beqs
g_{\rho\pi\pi}^2&\simeq&c_g\frac{M_{\rho}^2}{f_{\pi}^2}\,,
\eeqs
where $c_g>3$ is a model dependent numerical constant.
The choice of a small $\rho$ is equivalent to the choice of an enhanced
value for $f_{\pi}$, and hence a parametrical suppression
of the coupling of the techni-mesons to the $W$ and $Z$ gauge bosons.
As a result, the feature allowing for a very light spectrum of excitations,
would also result in a suppression of the production and decay rate of
these states. But moderately small values of $\rho\sim 1/2$ 
would be sufficient anyhow to evade the bounds, without incurring
in this problem. 
A more accurate study of the decay rate and production 
mechanism at the LHC is needed in order to determine how 
easy (or difficult) it could be to directly detect these states,
but this is not a parametrically big problem.

There is a substantial difference in the $d=2$ case.
Here, the boundary term $C$ is not required. 
In principle, it is certainly possible to 
add by hand such a term, and hence suppress the contribution
to $\hat{S}$ in the very same way as done in the 
$d=1$ case. But this would be an ad hoc, unnecessary
additional ingredient, essentially equivalent to adding
a higher-dimensional operator in the four-dimensional dual,
which would produce 
a tree-level contribution to $\hat{S}$
fine-tuned in such a way
as to cancel the contribution coming from the heavy states.
It is hence more interesting to study the model 
without this parameter.

As explained at length in the previous section,
the limit in which $\mu_W^2L_1^w\gg 1$ leads to a 
big modification of the polarizations at small momenta, and 
as a consequence to a big modification of the spectrum of 
the axial states. In particular, the mass-squared of the 
standard-model  $Z$ and $W$ gauge bosons are found to scale linearly with the
gauge coupling in this regime. This is obviously not compatible 
with the standard-model predictions, as
 seen from  the precision parameters, that turn out to be ${\cal O}(1)$
quantities (see for instance Eq.~(\ref{largeS})). This regime is hence clearly incompatible with 
experimental data, implying that $\mu_W^2L_1^2\ll1$.

The expression for $\hat{S}$ in the phenomenologically acceptable
range for the parameters of the model is hence 
\beqs
\hat{S}&\simeq&\frac{1}{e}M_W^2L_1^2\,,
\eeqs
which can be translated in a bound for the 
lightest techni-$\rho$ mass
\beqs
M_{\rho^0}&>&k\,(880 {\rm GeV})\,,
\eeqs
where $k$ is the same, $\varepsilon$-dependent constant 
of the $d=1$ case.
Depending on $\varepsilon$ this means
\beqs
M_{\rho^0}&>&(2\,-\,4) {\rm TeV}\,.
\eeqs
This result is in the upper limit of reach at the LHC.
If the most pessimist bound is assumed, it is difficult to believe that
the LHC signal for the new states could be the appearance of a resonance in the spectrum,
and much more elaborate data analysis strategies would be needed.
The lowest end of the limit however is within LHC reach
and is obtained for moderately small values of $\varepsilon \sim 1/3$.
Again, it is instructive to look at the $g_{\rho\pi\pi}$ coupling:
\beqs
g_{\rho\pi\pi}^2&\simeq&c_g \frac{M_{\rho}^2}{f_{\pi}^2}\,\\
&\simeq&c_g\frac{\varepsilon^2 g_{\rho}^2M_{\rho}^2}{M_Z^2}\,\\
&\simeq&c_g(g_4^2+g_4^{\prime\,2})\frac{M_{\rho}^2}{M_Z^2}\,,
\eeqs
where here $M_{\rho}=2.4/L_1$. 
This is certainly a strong coupling, though a more accurate estimate
is needed. The experimental signal of this model is hence expected to be quite
clear: there is no parametric suppression neither of the production
nor decay rate of the spin-1 techni-meson excitations, and it 
should be possible to detect them at the LHC.
In particular, even if the most pessimistic bound is assumed,
it should be possible to collect a large number of 
techni-$\rho$ decay events from the tails of its (broad)
resonance.

The bound on the mass of the lightest techni-$\rho$ of few TeV is
in substantial agreement with analogous estimates done
by simple re-scaling of QCD. One might wonder what 
has been gained here with this long exercise.
At first sight, it seems that $\hat{S}$ is suppressed by the same
naif idea of pushing $M_{\rho}$ to large values.
This fact deserves a comment.

If an interpretation of the present results in terms
of a four-dimensional theory holds, it does so only as long
as loop corrections are parametrically suppressed, i.e.
only at large-$N_c$.
At large $N_c$, in a QCD-like theory  the parametric separation between $f_{\pi}$
and $M_{\rho}$ disappears, and at the same time the decay
constants $f_{\rho}$ become bigger, because controlled by the
effective coupling itself. Hence, the large-$N_c$ 
expectation, based on the results from dispersion relations, 
is that $\hat{S}$ has to be very big, because none of the mechanisms
for its suppression is available.
By contrast, what found here is that the parametric separation 
between $M_{\rho}$ and $f_{\pi}$ can be obtained through walking dynamics,
even at large-$N_c$.
On top of that, walking makes the excited states less sensitive
to symmetry-breaking effects, so that, while it is still true
that  the decay constants $f_{\rho}$ are big,
this effect is compensated by a high degree of cancellation between
axial-vector and vector contribution, which is absent in a QCD-like model.
This high degree of degeneracy among the towers of axial and vectorial
excitations and their broadness are 
the most striking signature distinguishing the present models
from traditional  small-$N_c$ QCD-inspired models.

Concluding this section, in both models with $d=1$ and $d=2$ 
there are significant regions of parameter space that could be tested
at the LHC and are compatible with precision electro-weak data.
In both cases, the signature would be the existence of a set of strongly 
interacting spin-1 states with masses in the TeV region, and mass splitting
far too tiny for direct resolution. The decay modes of the resulting 
resonance  would hence comprise both
even and odd parity channels.

%%%%%%%%%%%%%%%%%%%%%%%%%%%%%%%%%%%%%%%%%%%%%%%%%%%%%%%%%%%%%%%%%%%%%%
\section{The Effect of Walking Behavior.}
%%%%%%%%%%%%%%%%%%%%%%%%%%%%%%%%%%%%%%%%%%%%%%%%%%%%%%%%%%%%%%%%%%%%%%

In the models discussed here, experimental bounds are
evaded by assuming the existence of a substantial
hierarchy between the masses of the SM gauge bosons and those of
their excited states. This hierarchy might, in general,  result from two 
complementary effects. 

One is the hierarchy between the weak couplings of
the SM and the strong effective coupling of the new states. This effect 
is parameterized by $\varepsilon$. The computations developed here are
reliable only for values of $\varepsilon$ not too small, because tiny 
values would translate into a large coupling $g$, 
and the perturbative expansion in the five-dimensional gauge theory would not hold.
Large or ${\cal O}(1)$ values of $\varepsilon$ correspond to large-$N_c$,
small values to  small-$N_c$. Hence, this cannot be the main reason why 
$\hat{S}$ is small enough to evade the experimental constraints.

The major effect is produced by the parametric separation between the symmetry
breaking scale and the confinement scale. At large $N_c$ one would expect
this to be just an ${\cal O}(1)$ effect. The claim here
is that walking can explain the presence of such a
hierarchy at large $N_c$, and as a consequence the bounds
on $\hat{S}$ are evaded by having a large enough mass
of the $techni-\rho$, while at the same time walking produces
also a quasi-degenerate spectrum of techni-$\rho$ and techni-$a_1$ states,
that compensates for the largeness of the decay constants for the excited states.

In order to asses the meaning of this assumption, I focus on the
$d=2$ case, which is the cleanest and less model-dependent.
The whole study performed here is supposed to describe the effect of
conformal symmetry (and walking) in modifying 
 the behavior of a QCD-like model, and as a tool for
the computation of non-perturbative corrections to the tree-level estimates
of the precision parameter $\hat{S}$. It is hence instructive to
compare the $d=2$ results to those of a   perturbative $SU(N_c)$ QCD-like
technicolor model, and to the non-perturbative estimates obtained
in absence of large anomalous dimension.

The perturbative estimate of $\hat{S}$ in a $SU(N_c)$ technicolor model
with $N_d$ fermions in the fundamental of both $SU(N_c)$ and 
$SU(2)_L$ can be written as
\beqs
\hat{S}_p &=& \frac{\alpha}{4\sin^2\theta_W}\frac{N_cN_d}{6\pi}\,.
\eeqs
For comparison, in the models discussed here, $N_c$ can be extracted from the
high-energy behavior of the vector-vector correlator,
and is given by:

\beqs
N_cN_d&=&\frac{12\pi^2}{g_\rho^2}\,\\
&=&\frac{12\pi^2\varepsilon^2}{g_4^2}\,\\
&=&\frac{3\pi \varepsilon^2\sin^2\theta_W}{\alpha}\,
\eeqs
and hence the perturbative estimate would be
\beqs
\label{perturbative}
\hat{S}_p&=&\frac{\varepsilon^2}{8}\,.
\eeqs
This estimate is in general agreement with what expected in a QCD-like 
technicolor model, since in this case $\varepsilon\sim g_4/g_{\rho} \sim 1/8$
can be estimated from the ratio of the mass of the the $\rho$ meson and 
the decay constant of the pion, though in this range of $\varepsilon$ the computations
performed here are not reliable and hence this comparison should not
be taken too literally.

The estimate in Eq.(\ref{perturbative}) is, obviously, a 
gigantic departure from what computed here, namely
\beqs
\label{nonperturbative}
\hat{S}&\simeq&\frac{\varepsilon^2}{2e}\mu_W^4L_1^4\,,
\eeqs
in which an additional suppression factor
is  coming from the parametric separation between the
symmetry breaking mass term for the spin-1 excitations 
 and the confinement scale $L_1$.
 Remember that, for $L_0\simeq L$, 
 \beqs
 \mu_W^4L_1^4&=&\frac{g^2}{4}\frac{\mbox{v}_0^2}{L^2}L_1^4\,
 =\,\frac{g^2}{4}\frac{\mbox{v}_1^2L_0^4}{L^2}\,
 \simeq\,\frac{g^2}{4}\mbox{v}_1^2L_0^2\,,
 \eeqs
 is the (dimensionless) 
 product of the coefficient of the symmetry-breaking term in the bulk equations
 for the charged gauge bosons, times the fourth power of the confinement scale.
 
The claim of this paper is that this suppression factor is natural,
and is precisely the effect of walking, which makes the 
non-perturbative estimate of $\hat{S}$ deviate by
orders of magnitude in respect to the perturbative estimates.
This claim seems to contradict analogous studies in the
literature, and hence it is worth comparing to the $d=3$ case.

In~\cite{pomarol}, the $d=3$ case is solved numerically. Imposing the 
experimental constraints that $M_{\rho}\simeq 770$ MeV,
$f_{\rho}\simeq 140$ MeV and $M_{a_1}\simeq 1230$ MeV,
the authors of~\cite{pomarol} find the predictions $f_{\pi}\simeq 87$ MeV
and $f_{a_1}\simeq 160$ MeV. These results are obtained
for a value of their symmetry-breaking parameter $\xi \simeq 4$,
while for small values of $\xi$, $f_{\pi}$ turns out to be parametrically small,
with $M_{\rho}\sim M_{a_1}$. For such large values of $\xi$,
the authors estimate $\hat{S}$, finding a modest dependence on the 
scaling dimension $d$.

For comparison, using of the exact expression for $\Sigma_{AA}$
in the $d=2$ case discussed here, after imposing
the same constraints on $M_{\rho}$ and $f_{\rho}$,
and requiring $f_{\pi}=87$ MeV,
I obtain the predictions for 
$M_{a_1} \simeq 1060$ MeV and $f_{a_1}\simeq 135$ MeV.
In particular, the ratio $f_{a_1}^2/M_{a_1}^2$ is substantially unchanged,
and accordingly $d$ has no significant effect on the estimate of $\hat{S}$.
The effect of the anomalous dimension is
a reduction of  the sensitivity to symmetry breaking of the excited modes
of the theory, in respect to the heavy modes.
However, this effect is not dramatic in this case,
consistently with what  pointed out in~\cite{pomarol}.

While the results in~\cite{pomarol} are consistent,
up to ${\cal O}(1)$ factors, with the perturbative effects,
this does not mean that non-perturbative effects are always small.
The reason for the smallness of the effect of changing $d$ 
observed in~\cite{pomarol}
is that the choice of parameters
giving the QCD-like spectrum
corresponds to the choice $\mu_W^4L_1^4\simeq 10$, and hence
no additional parametric suppression on $\hat{S}$ is present,
but, as also computed here in Eq.(\ref{largeS}),  $\hat{S}$ is
just proportional to $\varepsilon^2$ via a ${\cal O}(1)$ parameter.

The real question is then: are small values of $\mu_W^4L_1^4$
to be considered as natural, or is this just another 
way of recasting in a different language an old fine-tuning problem?
In order to answer this question, one has to associate the 
parameter $\mu_W^4L_1^4$ with the physical scales 
relevant for the system discussed here.

Looking back at the
definitions given in the setup, working at finite $L_0$,  before  regularization
and renormalization.
The SM fields, and hence the SM currents, are localized 
at the UV brane, where the symmetry breaking effects
are controlled by  $\mbox{v}_0$. This is the parameter that 
controls the magnitude of $G_F$, the Fermi constant.
The symmetry breaking effects experienced on the
IR-brane, i.e. for modes localized near the  IR brane, such as the first 
techni-$\rho$ excitations,
are rescaled as
\beqs
\mbox{v}_1&\sim&\left(\frac{L_1}{L_0}\right)^d \mbox{v}_0\,.
\eeqs
This is the parameter that controls $\hat{S}$.
If the technicolor dynamics consists of a large-$N_c$ 
 theory, with
 a quasi-conformal energy window
between  $L_0$ and $L_1$, but no anomalous dimensions,
the perturbative estimate of $\hat{S}$ should hold, and
hence the perturbative result should be consistent with a scaling
$\mbox{v}_1\sim (L_1/L_0)^3 \mbox{v}_0$.
If, on the contrary,  there is an anomalous dimension,  for the 
same value of $G_F$ and confinement scale $L_1$ one expects
the symmetry breaking
effects at the IR brane to be suppressed in respect to the perturbative
estimate, due to the different scaling of $\mbox{v}(z)$.
As a result
\beqs
\hat{S}&\sim&\left(\frac{L_0}{L_1}\right)^{6-2d}\,\hat{S}_p\,,
\eeqs
and in particular for the $d=2$ case one expects
a suppression 
\beqs
\label{scaling}
\hat{S}&\sim&\frac{L_0^2}{L_1^2}\hat{S}_p\,,
\eeqs
in respect to the perturbative estimate.

If these scaling arguments are correct, by comparing
Eq.~(\ref{perturbative}) and Eq.~(\ref{nonperturbative})
with Eq.~(\ref{scaling}) one is lead to conclude that
\beqs
\mu_W^4L_1^4 &\sim& \left(\frac{L_0}{L_1}\right)^2\,.
\eeqs

It remains to be proven that the very existence of 
a substantial hierarchy between $L_0$ and $L_1$ 
implies the smallness of $\mu_W^4L_1^4$.
The UV cut-off  $L_0$ corresponds to the $\Lambda_{ETC}$ scale,
at which the behavior of the theory changes, and up to
which the theory is (quasi) conformal. This is the scale at which 
higher-order interactions become important.
The IR cut-off is the technicolor scale $\Lambda_{TC}\sim 1/L_1$, 
at which the theory 
confines and a symmetry-breaking condensate is formed.
A hypothetical particle whose life
be confined on the IR brane, 
cannot experience other energy scales than
the one fixed by $L_1$, because conformal symmetry
is screening the effect of the UV cut-off.
It is unreasonable to expect the symmetry breaking condensate 
to form at a scale parametrically
higher than the confinement scale itself.
It is hence reasonable to write
\beqs
g\, \mbox{v}_1 &\propto& \frac{1}{L_1}\,,
\eeqs
with some proportionality factor that grows with the 
effective coupling, and hence cannot be more than a ${\cal O}(1)$ coefficient.
For this reason, in the limit in which $L_0\sim L$:
\beqs
\mu_W^4L_1^4 & \propto & \left(\frac{L_0}{L_1}\right)^2\,,
\eeqs
up to a coefficient that grows with the effective coupling
$g^2/L$. This is what desired.

Again, one could argue that I just transferred the fine-tuning problem of the smallness of 
$\hat{S}$, first into the fine-tuning problem of the smallness of $\mu_WL_1$, and 
then again in the fine-tuning problem of the hierarchy between $L_0$ and $L_1$, 
which rather than
a solution might look like a mere rewriting of the problem in a different language.
The point is that this hierarchy is natural, because
the scales $L_0$ and $L_1$, together with their separation, are
generated dynamically, and stabilized by conformal symmetry.
This deserves some more comments, in order to highlight the 
intrinsic limitations of this idea.

From the five-dimensional point of view, the existence of a 
hierarchy between $L_0$ and $L_1$ demands for a five-dimensional 
stabilization mechanism. The presence of a bulk scalar field
with non-trivial profile can itself provide such stabilization,
in the spirit of the Goldberger-Wise mechanism.
In order to produce an exponential separation of scales,
this mechanism would require a choice of $M^2L^2\simeq 0$,
very different from the choices discussed here.
Yet, for $d\simeq 2$, 
the modest factor of $L_0/L_1\sim1/10$,
required by  electro-weak precision measurements,
does not seem to pose severe problems.
This framework hence provides a 
natural suppression mechanism for $\hat{S}$.

If a four-dimensional interpretation holds, a generic $SU(N_c)$ gauge theory,
such as TC is supposed to be, is not going to exhibit  conformal behavior
at large coupling  in the IR energy region. 
Nevertheless, an accurate choice of the fermionic field 
content can suppress the beta-function coefficient, and the running of the gauge
coupling, below some symmetry breaking scale $L_0$. Conformal, strong-interacting,
gauge symmetries of this type
have been studied at length, both in the context of supersymmetric and
non-supersymmetric models.
The fact that at the UV-scale $L_0$ the coupling is already strong,
together with the fact that in real-world models exact conformality is 
an unreasonable assumption, implies that the scale $L_1$ at which the 
theory finally confines cannot be exponentially far away from $L_0$. 
But it is perfectly reasonable to think that 
models with $L_0/L_1\sim 1/10$ exist and are natural. 
And this is the size required
by the phenomenology discussed in the present paper.
As a result, walking behavior can naturally suppress the electro-weak precision 
parameters, that hence do not constrain the number $N_c$ of techni-colors
to be very small, as opposed to what indicated by perturbative estimates.

By contrast, using the arguments developed here,
one sees that the choice of large value of $\mu_W^4L_1^4\sim 10$,
dictated by the attempt to reproduce low-energy QCD data,
corresponds to a regime in which $L_0\sim L_1$, 
with a largish five-dimensional coupling $g$.
The effective coupling is large because QCD is a small-$N_c$ theory, 
and in this regime the computations performed here have large systematic uncertainties.
More important  is the observation that
 in QCD a conformal energy window in the IR (at large coupling)
does not exist, and the QCD coupling actually runs fast
in the energy region
around $.5 - 3$ GeV, instead of walking.
This enforces $L_0\simeq L_1$, 
and explains why changing $d$ does not affect significantly the results
for $\hat{S}$ in this class of models~\cite{pomarol} of QCD-like theories.
As a result, these models do not provide a  good description
of QCD in the interesting intermediate energy 
region just above the confinement scale, 
as is well illustrated by the fact that the Regge 
trajectories are not reproduced. General features of 
low-energy QCD  require a substantial departure
of the background from the pure $AdS_5$ (i.e. from conformal symmetry)
in the IR in order for a model to reproduce the data.

The toy models discussed here show that the presence of 
a sizable energy window in which the theory is conformal,
and in which the condensates have large, non-perturbative,
anomalous dimensions, provide a suppression mechanism
for $\hat{S}$ which is a viable alternative to those discussed for instance in~\cite{composite}
and~\cite{Sannino}. In the latter  class of models, the suppression is
achieved by enhancing the scale of strong dynamics
above the electro-weak scale, and by explaining 
electro-weak symmetry breaking with the VEV of a light
composite scalar emerging from the higher scale condensation,
in the spirit of composite Higgs and little Higgs models.
The main phenomenological distinction is that
in the models discussed in the present paper there is no light scalar,
while in this other class of models there are in general several
very light physical scalars, whose masses tend to be even too
small to satisfy the present experimental bounds from direct searches.

%%%%%%%%%%%%%%%%%%%%%%%%%%%%%%%%%%%%%%%%%%%%%%%%%%%%%%%%%%%%%%%%%%%%%%
\section{Conclusions.}
%%%%%%%%%%%%%%%%%%%%%%%%%%%%%%%%%%%%%%%%%%%%%%%%%%%%%%%%%%%%%%%%%%%%%%

The two examples of models discussed here share the 
advantages of being exactly solvable, and 
of assuming the presence of large anomalous dimensions
for the chiral condensate.
The accuracy of the computations performed is limited by
the fact that only the bilinear terms in the action
have been retained, and hence they can be used 
to describe a $SU(N_c)$  model only in the large $N_c$ limit.

From the model-building perspective.
At large $N_c$, the results show a huge parametric departure from 
the perturbative estimate, in the form of a suppression of $\hat{S}$
proportional to powers of  the ratio between the upper and lower energy scales between which
the theory is approximately conformal ({\it walks}) in the IR.
This suppression relies on the existence of a stabilization
mechanism between these two scales and on the existence of 
condensates with large anomalous dimensions,
both of which are natural in the context of the $AdS-CFT$ correspondence.
Walking is anyhow required in the construction
of semi-realistic models of  dynamical electro-weak symmetry breaking
by the requirement of providing large enough masses for the SM fermions,
while at the same time suppressing flavor-changing neutral current processes.
This indicates that 
electro-weak precision measurements do not constrain these models
to have a small number $N_c$ of techni-colors, but rather they confirm the 
fact that a non-perturbative, large anomalous dimension for the chiral condensate
and a significant regime of quasi-conformal behavior in the IR
are  necessary requirements
in the construction of  viable models.

On the phenomenological aspects.
This study indicates that  models of this class
are compatible with electro-weak precision 
data for a range of masses of the excited spin-1 states
that is testable at the LHC. This result is, for all practical purposes,
independent of the number of techni-colors of the underlying strong dynamics.
The level of degeneracy between the spectrum of techni-$\rho$ and techni-$a_1$
resonances provides a distinction between QCD-like technicolor models
with small $N_c$ and walking technicolor theories with large $N_c$. 
The practical feasibility of such a search is subject to the precise determination 
of the strength of the coupling of the new states to the SM currents and gauge bosons,
which requires further investigation, but there is no reason to expect
these couplings  to be suppressed. These resonances should be
quite  broad, so that a significant number of events should be detectable
even for masses in the upper limit
of energy reached at the LHC.

In the end, LHC experimental data will tell us wether these models have anything to do
with nature or not. What this study shows is that large-$N_c$ dynamical
electro-weak symmetry breaking models
are not ruled out by present indirect constraints from electro-weak precision
measurements, and have a very distinctive and clear experimental signature
that is  testable at the LHC.

%%%%%%%%%%%%%%%%%%%%%%%%%%%%%%%%%%%%%%%%%%%%%%%%%%%%%%%%%%%%%%%%%%%%%%
%% Acknowledgments %%%%%%%%%%%%%%%%%%%%%%%%%%%%%%%%%%%%%%%%%%%%%%%%%%%
%%%%%%%%%%%%%%%%%%%%%%%%%%%%%%%%%%%%%%%%%%%%%%%%%%%%%%%%%%%%%%%%%%%%%%
\vspace{1.0cm}
\begin{acknowledgments}
I would like to thank Thomas Appelquist, 
Aleksi Vuorinen,  Matthew Strassler, Dam Son,
Walter Goldberger,  Christopher Herzog and in particular 
Andreas Karch, for very useful discussions and comments.
This work is partially supported by the Department of Energy grant
DE-FG02-96ER40956.
\end{acknowledgments}

%%%%%%%%%%%%%%%%%%%%%%%%%%%%%%%%%%%%%%%%%%%%%%%%%%%%%%%%%%%%%%%%%%%%%%
%%%  Bibliography  %%%%%%%%%%%%%%%%%%%%%%%%%%%%%%%%%%%%%%%%%%%%%%%%%%%
%%%%%%%%%%%%%%%%%%%%%%%%%%%%%%%%%%%%%%%%%%%%%%%%%%%%%%%%%%%%%%%%%%%%%%

\end{document}